\documentclass[preprint,12pt,authoryear]{elsarticle}
\usepackage[T1]{fontenc}
\usepackage[utf8]{inputenc}
\usepackage{newtxtext,newtxmath}
\usepackage{microtype,amsmath,booktabs,array,tabularx,graphicx}
\usepackage{placeins}
\usepackage{xcolor}
\usepackage{xurl}
\usepackage[hidelinks,pdfencoding=auto]{hyperref}
\journal{Journal of Forecasting}
\biboptions{authoryear,round}
\newcolumntype{L}[1]{>{\raggedright\arraybackslash}p{#1}}
\newcolumntype{C}[1]{>{\centering\arraybackslash}p{#1}}

\newcommand{\dd}{\,\mathrm{d}}

\newcommand{\notes}[1]{\par\vspace{4pt}\begin{minipage}{\linewidth}\footnotesize #1\end{minipage}}
\newcommand{\gain}[3]{\begin{tabular}[t]{@{}c@{}}#1\\[-2pt]{\scriptsize[#2,\,#3]}\end{tabular}}

\newif\ifdraftnotes
\draftnotesfalse
\ifdraftnotes
  \newcommand{\todo}[1]{{\color{red}\footnotesize\sffamily[TODO #1]}}
  \newcommand{\tbd}[1][]{{\color{red}\textbf{??}\ifx\relax#1\relax\else\textsuperscript{\,(old:\,#1)}\fi}}
  \newcommand{\hyp}[1]{{\color{blue}#1}}
\else
  \newcommand{\todo}[1]{\PackageError{draftnotes}{An unresolved TODO remains}{Resolve it and delete the note.}}
  \newcommand{\tbd}[1][]{\PackageError{draftnotes}{A missing number (\string\tbd) remains}{Insert the computed value.}}
  \newcommand{\hyp}[1]{\PackageError{draftnotes}{An unverified sentence (\string\hyp) remains}{Rewrite it from the results and remove the wrapper.}#1}
\fi

\hypersetup{pdftitle={Scale or speed? When do path signatures improve volatility-regime forecasts?},pdfauthor={Yang Lyu; Kiseop Lee}}

\begin{document}
\begin{frontmatter}
\title{Scale or speed? When do path signatures improve volatility-regime forecasts?\tnoteref{run}}
\tnotetext[run]{Running title: Signatures and volatility forecasts}

\author[purdue]{Yang Lyu}
\ead{lyu143@purdue.edu}
\author[purdue]{Kiseop Lee\corref{cor1}}
\ead{kiseop@purdue.edu}
\cortext[cor1]{Corresponding author.}

\affiliation[purdue]{organization={Department of Statistics, Purdue University},
            addressline={250 N. University Street},
            city={West Lafayette},
            postcode={47907},
            state={IN},
            country={USA}}

\begin{abstract}
We study when path signatures, increasingly used as model-free features of return paths, improve out-of-sample forecasts of volatility regimes. In daily United States and developed-market equity returns from 2005 to 2025, machine-learning classifiers do not beat a heterogeneous autoregressive (HAR) regression of realized volatility, although signatures improve the classifiers' accuracy and probability forecasts in the United States. Adding raw signatures to the HAR regression itself improves United States forecasts of next-month volatility terciles, raising balanced accuracy by 2.7 percentage points and lowering the ranked probability score after multiple-testing correction, whereas adding conventional volatility features does not; outside the United States no gain survives correction. Controlled Heston experiments explain the pattern: signatures add modestly when regimes differ in variance level, which realized volatility already captures, and more when regimes share one stationary variance law but differ in mean-reversion speed. Better regime identification does not, however, deliver earlier regime-change signals, and no accuracy gain translates into a resolved Sharpe-ratio gain in a volatility-managed portfolio. Signatures thus add information about how volatility evolves, which realized-volatility predictors miss.
\end{abstract}


\end{frontmatter}

\section{Introduction}\label{sec:intro}
Forecasts of volatility regimes feed risk budgets, volatility targeting and asset allocation, and regime-switching models are a standard way to produce them \citep{hamilton1989,ang2012}. Statistical jump models, which penalize state changes \citep{nystrup2020}, now supply regime signals for downside-risk control \citep{shu2024}, and machine-learning representations of return paths are proposed as model-free predictors. Path signatures, which summarize a path through its iterated integrals \citep{lyons2014,chevyrev2016}, have been applied to market-regime classification and clustering \citep{bilokon2021,horvath2024}, online regime detection \citep{issa2023}, identification of volatility models \citep{bures2026} and quickest detection \citep{wang2026}. Whether they improve out-of-sample regime forecasts beyond the volatility predictors that forecasters already use, such as the heterogeneous autoregressive (HAR) averages of realized variance \citep{corsi2009}, is less clear.

Reported signature gains are hard to interpret for three reasons. First, regimes that differ in variance level are already separated by realized volatility, so accuracy in such settings reveals little about the information that signatures add. Second, gains are usually reported for one learner and one baseline, although representation and learner interact: continuous functionals of a path are approximately linear in its signature \citep{lyons2014}, but not in its log-signature. Third, identifying a regime once it is established differs from signalling its onset, and a forecaster who acts on regime changes needs both.

We therefore ask when signatures improve volatility-regime forecasts, and why. Our main evidence is an out-of-sample forecasting study: we forecast the tercile of next-month realized volatility in daily United States and developed-market equity returns from 2005 to 2025 and evaluate the forecasts by balanced accuracy and by the ranked probability score. We compare machine-learning classifiers built on a compact and a rich set of conventional predictors, each with and without signatures, with a hidden Markov model and with conventional benchmarks, including persistence and a HAR regression, and we ask whether signatures improve the HAR regression itself. We also translate the forecasts into a volatility-managed portfolio \citep{moreira2017}. To explain the results, we use two controlled Heston designs: in one, regimes differ in variance scale; in the other, they share the same stationary variance law but differ in speed. Finally, we ask whether better regime identification also yields earlier regime-change signals at matched false-alarm rates, against a cumulative-sum benchmark \citep{page1954}.

Four findings emerge. First, the conventional benchmarks are hard to beat: the HAR regression forecasts next-month terciles as accurately as any classifier, with or without signatures, and persistence of the current tercile is not far behind. Signatures nevertheless improve the classifiers built on the compact predictor set in the United States, in balanced accuracy and in the ranked probability score. Second, signatures also improve the HAR regression: in the United States, adding raw signatures raises balanced accuracy by 2.7 percentage points and lowers the ranked probability score, both resolved after Holm correction, whereas adding the conventional features to the same regression changes nothing; outside the United States no gain survives correction. Third, the controlled designs show why: signatures carry information about how regimes evolve rather than how volatile they are, their gain is larger when regimes differ in speed than in scale in eleven of twelve augmentation contrasts, and it is largest when raw signatures feed a linear learner. Fourth, the gains have limited practical reach: better identification of established regimes does not produce earlier regime-change signals, and no accuracy gain translates into a resolved Sharpe-ratio improvement in a volatility-managed portfolio. The forecasting value of signatures is therefore conditional in a checkable way: on whether regimes differ in level or in dynamics, on the benchmark they augment, on the learner, and on whether the forecaster needs established regimes or early signals of change.

Section~\ref{sec:framework} describes the forecasting framework, Section~\ref{sec:forecasts} reports the out-of-sample forecasts, Section~\ref{sec:mechanism} explains them with controlled experiments, and Section~\ref{sec:conclusion} concludes. \ref{app:detection} reports the detection experiment.

\section{Forecasting framework}\label{sec:framework}
\subsection{Data and forecasting target}\label{sec:data}
We use daily value-weighted market returns for the United States and for developed markets excluding the United States from the Kenneth French Data Library \citep{french2026}, July 1990 to December 2025. Log returns are computed from the market excess return plus the risk-free rate. On each day $t$, the forecast target is the tercile of annualized realized volatility over the next 21 trading days, so the three classes are low-, medium- and high-volatility months ahead; tercile cut-points are estimated on training data only. Forecasts are issued daily from information available on day $t$.

\subsection{Predictors}\label{sec:predictors}
Each forecast uses the most recent $W=50$ daily returns. For the window ending on day $t$, let $r_a=\log S_{t-W+a}-\log S_{t-W+a-1}$, $a=1,\ldots,W$, denote its log returns in chronological order, and let
\begin{equation}
 x_j=\left(j/W,\ \sum_{a=1}^{j}r_a,\ \sum_{a=1}^{j}|r_a|\right),
 \qquad j=0,\ldots,W.
 \label{eq:path}
\end{equation}
be the origin-anchored, piecewise-linear path of time, cumulative return and cumulative absolute return. The time channel makes the representation sensitive to when variation occurs within the window. For example, the level-2 signature contains the L\'evy area between time and cumulative absolute return, which records whether large absolute returns occur early or late in the window.

The compact conventional predictor set, denoted Statistics, has 11 features: volatility over the full and short windows, mean absolute return, downside and upside semivariance, skewness, excess kurtosis, lag-one autocorrelations of returns, absolute returns and squared returns, and maximum log drawdown. The rich set, B3, contains all 11 Statistics features together with 29 additional features: return quantiles, autocorrelations at lags 2, 5 and 10, variance ratios, spectral entropy and frequency-band powers, realized variance and quarticity, bipower variation, a jump fraction, and sign-change and turning-point rates. B3 is therefore a strictly richer conventional baseline. In the forecasting study both sets also include averages of squared returns over 1, 5 and 22 days, as in the HAR model \citep{corsi2009}, giving 14 and 43 conventional features.

The level-3 log-signature of path~\eqref{eq:path} has 14 coordinates and the level-3 raw signature 39; both are computed with \texttt{iisignature} \citep{reizenstein2020} and appended to Statistics or B3. As a parametric benchmark, we fit a three-state Gaussian hidden Markov model to the training returns \citep{hamilton1989}; its filtered state probabilities, given all returns up to day $t$, form a further predictor block.

\subsection{Forecasting models}\label{sec:models}
Each predictor set is combined with three learners: \(\ell_2\)-regularized logistic regression, random forests and histogram gradient boosting, each with a prespecified three-setting hyperparameter grid (Supporting Information, Section S2).

Three conventional benchmarks require no learner. Persistence forecasts next month's tercile by the tercile of realized volatility over the last 21 days. A transition benchmark turns persistence into a probability forecast, using the training-sample frequencies of next-month terciles given the current tercile, with add-one smoothing. The HAR benchmark regresses log next-month realized volatility on the logs of realized volatility over the last 1, 5 and 22 days \citep{corsi2009} and converts its Gaussian predictive distribution into tercile probabilities; because some days have zero returns, the one-day component is floored at its training 1\% quantile.

To ask whether signatures add to this benchmark, we augment the HAR regression with standardized extra predictors: the 14 log-signature coordinates, the 39 raw signature coordinates or, as a conventional control, the 11 Statistics features. Only the extra coefficients are ridge-penalized. The penalty is chosen on the validation years by the ranked probability score from a grid that includes an infinite penalty, which returns the plain HAR forecast, and the probabilities are formed as for the HAR benchmark. This comparison was designed after the benchmark results were known but fixed before any augmented forecast was scored.

\subsection{Walk-forward estimation and forecast evaluation}\label{sec:evaluation}
Models are refitted each year on an expanding window, with a 21-day gap before each test year so that no training label overlaps the test period, and each representation's learner, hyperparameters and ridge penalty are selected on the last three years of the training window. The out-of-sample period is 2005--2025, giving 21 annual test years, and no test observation is used for any selection.

Forecasts are evaluated in two ways. Balanced accuracy (BA), the unweighted mean of the three class recalls, scores the most probable tercile and is insensitive to the unequal frequency of terciles within a test period. The ranked probability score (RPS; \citealp{epstein1969}) scores the full probability forecast: it sums the squared differences between cumulative forecast and outcome probabilities over the ordered terciles, and we report it multiplied by 100, so that lower values are better. The logarithmic and Brier scores, both strictly proper \citep{gneiting2007}, point in the same direction in every contrast and are reported in the Supporting Information (Table S13). Differences in BA are assessed with 95\% stationary-bootstrap intervals based on a mean block length of 21 days \citep{politis1994}, and differences in scores with Diebold--Mariano tests \citep{diebold1995} using a Bartlett HAC variance with bandwidth 21 \citep{newey1987}, which accounts for the overlap of 21-day targets. The familywise error rate is controlled with the Holm correction within each family of contrasts: the eight signature augmentations of the classifiers (two markets, two baselines and two signature types) and the six augmentations of the HAR regression (two markets and three predictor sets).

\subsection{Economic evaluation}\label{sec:economic}
To assess the economic value of the forecasts, we scale next-day market exposure by the inverse of forecast variance, $w_t=c/\hat\sigma_t^2$, where $\hat\sigma_t^2=\sum_k\hat p_{t,k}\bar v_k$ combines the predicted regime probabilities $\hat p_{t,k}$ with training-sample regime variances $\bar v_k$, $c$ matches the training-sample volatility of the market, and leverage is capped at two \citep{moreira2017}. We evaluate two signature rules, fixed before the market data were analyzed: Statistics plus log-signatures and B3 plus raw signatures, the headline augmentations of the controlled experiments in Section~\ref{sec:mechanism}. The Statistics and HMM rules serve as conventional benchmarks, and Sharpe-ratio differences are tested with the method of \citet{ledoit2008}. The same rule is later applied to the plain and augmented HAR forecasts, with the plain HAR rule as the benchmark. The exercise asks whether forecast differences are economically visible; it is not a proposed trading strategy.

\section{Out-of-sample forecasts}\label{sec:forecasts}
\subsection{Benchmarks and classifiers}\label{sec:classifiers}
Table~\ref{tab:market} compares the classifier forecasts with the benchmarks. The benchmarks are hard to beat. Persistence of the current tercile already reaches 55.81\% balanced accuracy in the United States and 56.82\% elsewhere, and the HAR regression reaches 56.22\% and 57.51\%, with the lowest RPS among the benchmarks. No classifier beats HAR in either market: the most accurate United States classifier, Statistics augmented with raw signatures, exceeds HAR by 1.71 points [$-1.67$, 5.11] in balanced accuracy, and its RPS differs from that of HAR by 0.15 [$-0.88$, 1.14]. Outside the United States every classifier falls short of HAR in balanced accuracy.

Within the classifiers, signatures improve the compact predictor set in the United States. Adding raw signatures to Statistics raises balanced accuracy by 3.68 points [1.36, 6.03] and lowers the RPS by 1.72 [0.74, 2.62], and log-signatures lower the RPS by 1.52 [0.40, 2.68]; all three contrasts survive Holm correction. The raw coordinate system again adds more than the log coordinates, as the controlled experiments of Section~\ref{sec:mechanism} lead one to expect. Outside the United States no signature augmentation of a classifier is resolved in either measure.

\begin{figure}[htbp]
\centering
\includegraphics[width=\linewidth]{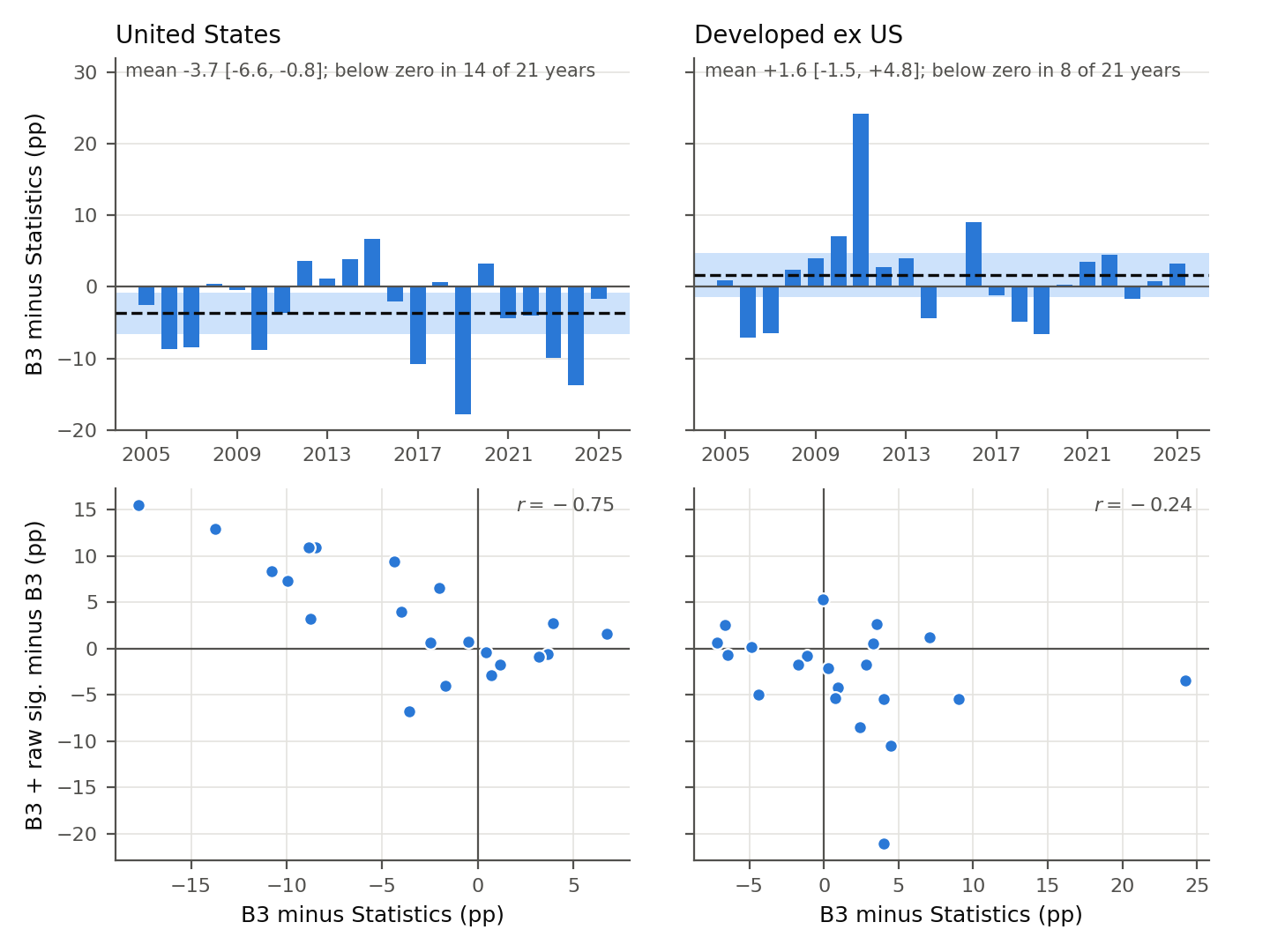}
\caption{Out-of-sample balanced accuracy of B3 relative to Statistics by test year, 2005--2025, in the United States (left) and in developed markets excluding the United States (right). Top: yearly difference in percentage points, with its mean (dashed line) and 95\% Student-$t$ interval over years (band). Bottom: yearly gain from adding raw signatures to B3 against the yearly B3-minus-Statistics difference, with their correlation $r$. Each representation uses its validation-selected learner in each year.}
\label{fig:b3year}
\end{figure}

The rich predictor set requires a different reading. B3 attains only 50.26\% in the United States, almost four points below the compact set that it contains, and falls below Statistics in 14 of the 21 test years (Figure~\ref{fig:b3year}). Adding log- and raw signatures raises its balanced accuracy by 3.60 [1.30, 5.98] and 3.72 [0.94, 6.31] points, both resolved after Holm correction, and log-signatures also lower its RPS by 1.04 [0.40, 1.64]; yet the augmented forecasts, at 53.87\% and 53.98\%, remain below Statistics alone. These contrasts measure how much of a degraded baseline the signatures recover, not an improvement over the best forecast. Year by year, the gain from adding raw signatures to B3 is largest when B3 falls furthest below Statistics (correlation $-0.75$). Outside the United States B3 shows no such shortfall: it lies above Statistics (55.59\% against 54.77\%) and falls below it in only 8 of the 21 test years, and in neither market does its year-to-year variability differ detectably from that of Statistics (Supporting Information, Section S7).

\begin{table}[htbp]
\centering
\caption{Out-of-sample forecasts of next-month volatility terciles: benchmarks and classifiers, 2005--2025.}
\label{tab:market}
\footnotesize
\begin{tabular*}{\linewidth}{@{\extracolsep{\fill}}lcccc@{}}
\toprule
& \multicolumn{2}{c}{United States} & \multicolumn{2}{c}{Developed ex US}\\
\cmidrule(lr){2-3}\cmidrule(l){4-5}
& BA (\%) & RPS & BA (\%) & RPS\\
\midrule
\multicolumn{5}{@{}l}{\emph{Panel A. Forecast accuracy}}\\
Persistence & 55.81 & 25.29 & 56.82 & 23.96 \\
Transition & 55.81 & 17.53 & 56.82 & 17.27 \\
HAR & 56.22 & 16.35 & 57.51 & 16.04 \\
HMM probability block & 56.10 & 16.58 & 56.00 & 16.47 \\
Statistics & 54.25 & 18.22 & 54.77 & 15.88 \\
Statistics + log-sig. & 56.38 & 16.70 & 56.09 & 16.40 \\
Statistics + raw sig. & 57.92 & 16.50 & 54.28 & 16.74 \\
B3 & 50.26 & 18.18 & 55.59 & 17.39 \\
B3 + log-sig. & 53.87 & 17.14 & 54.46 & 17.03 \\
B3 + raw sig. & 53.98 & 18.18 & 52.40 & 17.92 \\
\midrule
\multicolumn{5}{@{}l}{\emph{Panel B. Signature augmentation: BA gain and RPS reduction [95\% CI]}}\\
Statistics $\to$ + log-sig. & \gain{+2.13}{$-$0.68}{4.88} & \gain{\textbf{+1.52}}{0.40}{2.68} & \gain{+1.32}{$-$1.19}{3.84} & \gain{$-$0.51}{$-$1.17}{0.25} \\[8pt]
Statistics $\to$ + raw sig. & \gain{\textbf{+3.68}}{1.36}{6.03} & \gain{\textbf{+1.72}}{0.74}{2.62} & \gain{$-$0.49}{$-$3.70}{2.88} & \gain{$-$0.86}{$-$1.85}{0.22} \\[8pt]
B3 $\to$ + log-sig. & \gain{\textbf{+3.60}}{1.30}{5.98} & \gain{\textbf{+1.04}}{0.40}{1.64} & \gain{$-$1.13}{$-$3.33}{0.99} & \gain{+0.36}{$-$0.32}{0.95} \\[8pt]
B3 $\to$ + raw sig. & \gain{\textbf{+3.72}}{0.94}{6.31} & \gain{$-$0.01}{$-$0.94}{0.97} & \gain{$-$3.19}{$-$5.82}{$-$0.53} & \gain{$-$0.53}{$-$1.39}{0.44} \\
\bottomrule
\end{tabular*}
\notes{BA is the balanced accuracy of the most probable tercile; RPS is the ranked probability score multiplied by 100 (lower is better). Persistence forecasts the tercile of realized volatility over the last 21 days; Transition uses the training frequencies of next-month terciles given the current tercile; HAR regresses log next-month realized volatility on log realized volatility over the last 1, 5 and 22 days, with Gaussian residuals. Classifier rows use the validation-selected learner; Statistics and B3 include HAR-type averages of squared returns. Panel B reports the change from adding signatures to the base block, with 95\% stationary-bootstrap intervals (mean block length 21 days, 2,000 resamples); positive values are improvements. Bold marks a contrast resolved at the 5\% level after Holm correction within the eight contrasts of each measure (BA: bootstrap test; RPS: Diebold--Mariano test with Bartlett HAC variance, bandwidth 21). Data: Kenneth French Data Library.}
\end{table}

\subsection{Signatures in the HAR regression}\label{sec:har}
Table~\ref{tab:harsig} asks whether signatures add to the strongest benchmark. In the United States, augmenting the HAR regression with raw signatures raises balanced accuracy from 56.22\% to 58.92\%, a gain of 2.71 points [0.54, 4.73], and lowers the RPS from 16.35 to 15.33, a reduction of 1.02 [0.36, 1.66]; both survive Holm correction across the six HAR contrasts, as does the reduction in the Brier score. These are the most accurate forecasts in the study by either criterion. Log-signatures give smaller gains that are resolved only nominally. The conventional control fails: adding the 11 Statistics features to the same regression leaves balanced accuracy and the RPS essentially unchanged, at $-0.44$ points and 0.18. The gain therefore reflects information specific to the signature rather than additional predictors as such. The validation years return the plain HAR forecast for the signature augmentations in at most one of the 21 test years per market, but in three years for the Statistics control. Outside the United States, log-signatures lower the RPS by 0.47 [0.06, 0.88], which is not resolved after Holm correction, and no augmentation improves balanced accuracy.

\begin{table}[htbp]
\centering
\caption{HAR forecasts augmented with signatures, 2005--2025.}
\label{tab:harsig}
\footnotesize
\begin{tabular*}{\linewidth}{@{\extracolsep{\fill}}lcccc@{}}
\toprule
Model & BA (\%) & RPS & BA gain [95\% CI] & RPS reduction [95\% CI]\\
\midrule
\multicolumn{5}{@{}l}{\emph{Panel A. United States}}\\
HAR & 56.22 & 16.35 & -- & -- \\
HAR + Statistics & 55.77 & 16.17 & \gain{$-$0.44}{$-$2.14}{1.29} & \gain{+0.18}{$-$0.31}{0.69} \\[8pt]
HAR + log-sig. & 58.49 & 15.53 & \gain{+2.27}{0.39}{4.13} & \gain{+0.82}{0.15}{1.58} \\[8pt]
HAR + raw sig. & 58.92 & 15.33 & \gain{\textbf{+2.71}}{0.54}{4.73} & \gain{\textbf{+1.02}}{0.36}{1.66} \\[8pt]
\addlinespace
\multicolumn{5}{@{}l}{\emph{Panel B. Developed markets excluding the United States}}\\
HAR & 57.51 & 16.04 & -- & -- \\
HAR + Statistics & 56.19 & 15.92 & \gain{$-$1.32}{$-$2.92}{0.33} & \gain{+0.13}{$-$0.32}{0.60} \\[8pt]
HAR + log-sig. & 57.19 & 15.57 & \gain{$-$0.32}{$-$1.95}{1.42} & \gain{+0.47}{0.06}{0.88} \\[8pt]
HAR + raw sig. & 56.17 & 15.72 & \gain{$-$1.34}{$-$3.69}{1.15} & \gain{+0.33}{$-$0.30}{0.95} \\
\bottomrule
\end{tabular*}
\notes{The HAR regression of Table~\ref{tab:market} is augmented with standardized extra predictors: the 11 Statistics features (conventional control), the 14 level-3 log-signature coordinates or the 39 level-3 raw signature coordinates. Only the extra coefficients are ridge-penalized; the penalty is chosen on the validation years by the RPS from a grid that includes an infinite penalty (the plain HAR forecast). RPS is multiplied by 100. Gains are relative to HAR, with 95\% stationary-bootstrap intervals; bold marks a contrast resolved at the 5\% level after Holm correction within the six contrasts of each measure. The comparison was designed after the benchmark results were known and fixed before any augmented forecast was scored.}
\end{table}

\subsection{Economic value}\label{sec:econvalue}
Table~\ref{tab:econ} evaluates the two prespecified signature rules economically. Point estimates favor the signature rule in three of the four comparisons: in the United States, B3 plus raw signatures attains a Sharpe ratio of 0.609 against 0.566 for the Statistics rule ($p=0.45$) and Statistics plus log-signatures attains 0.469 ($p=0.32$); in developed markets excluding the United States the corresponding ratios are 0.525 and 0.472 against 0.410 ($p=0.10$ and $p=0.13$). No difference is resolved, but the test has little power against Sharpe-ratio differences of this magnitude in a 21-year daily sample, so the evidence neither establishes nor excludes an economic benefit. Managed rules reduce maximum drawdown relative to buy-and-hold in both markets, but this pattern is shared by the conventional rules and primarily reflects inverse-variance scaling; signature augmentation does not consistently reduce drawdown further. The two signature rules that were not prespecified are reported in the Supporting Information (Section S7); none of the ten Sharpe-ratio comparisons across the two markets is resolved after Holm correction. The same exposure rule applied to the HAR forecasts gives a similar picture (Table~\ref{tab:hareco}). The plain HAR rule attains a Sharpe ratio of 0.598 in the United States and 0.339 elsewhere, and the rules based on the augmented HAR regressions do not improve on it: in the United States, HAR plus raw signatures, the most accurate forecast in the study, attains 0.550 ($p=0.24$), and outside the United States the largest difference, 0.392 for HAR plus log-signatures ($p=0.02$), is not resolved after Holm correction across the six comparisons. The gains in forecast accuracy are therefore not visible in the Sharpe ratio of a volatility-managed portfolio. This evaluation was added after the augmented forecasts had been scored.

Two questions follow: why signatures add to the HAR regression when conventional features do not, and why their measured value in the classifiers depends so strongly on the baseline. The controlled experiments of Section~\ref{sec:mechanism} address both.

\begin{table}[htbp]
\centering
\caption{Volatility-managed market exposure, 2005--2025.}
\label{tab:econ}
\begin{tabular*}{\textwidth}{@{\extracolsep{\fill}}lcccccc@{}}
\toprule
Exposure rule & Mean & Vol. & Sharpe & Max DD & Turnover & $p$ \\
\midrule
\multicolumn{7}{@{}l}{\textit{Panel A. United States}} \\
Buy-and-hold          & 10.60 & 19.43 & 0.546 & $-55.8$ & --    & --   \\
HMM probability block & 10.17 & 17.35 & 0.586 & $-39.9$ & 0.117 & 0.78 \\
Statistics            &  9.53 & 16.85 & 0.566 & $-33.7$ & 0.067 & --   \\
Statistics + log-sig. &  8.74 & 18.65 & 0.469 & $-48.3$ & 0.069 & 0.32 \\
B3 + raw sig.         & 10.55 & 17.33 & 0.609 & $-37.2$ & 0.061 & 0.45 \\
\addlinespace
\multicolumn{7}{@{}l}{\textit{Panel B. Developed markets excluding the United States}} \\
Buy-and-hold          &  5.80 & 15.73 & 0.369 & $-60.2$ & --    & --   \\
HMM probability block &  6.50 & 15.75 & 0.413 & $-48.0$ & 0.085 & 0.95 \\
Statistics            &  6.23 & 15.21 & 0.410 & $-46.6$ & 0.055 & --   \\
Statistics + log-sig. &  7.29 & 15.45 & 0.472 & $-45.1$ & 0.060 & 0.13 \\
B3 + raw sig.         &  8.42 & 16.02 & 0.525 & $-51.2$ & 0.081 & 0.10 \\
\bottomrule
\end{tabular*}
\notes{Daily excess returns of the value-weighted market in each region. Exposure on day $t+1$ is $w_t=c/\hat{\sigma}_t^2$, capped at two, where $\hat{\sigma}_t^2$ is the variance implied by predicted regime probabilities and $c$ equates strategy and market volatility over the training sample. Mean and Vol.\ are annualized and reported in percent; Max DD is maximum drawdown in percent; Turnover is average daily $|w_t-w_{t-1}|$. The $p$-value tests equality of the rule's Sharpe ratio and that of the Statistics rule using the method of \citet{ledoit2008}.}
\end{table}

\begin{table}[htbp]
\centering
\caption{Volatility-managed market exposure based on the HAR forecasts, 2005--2025.}
\label{tab:hareco}
\begin{tabular*}{\textwidth}{@{\extracolsep{\fill}}lcccccc@{}}
\toprule
Exposure rule & Mean & Vol. & Sharpe & Max DD & Turnover & $p$ \\
\midrule
\multicolumn{7}{@{}l}{\textit{Panel A. United States}} \\
HAR & 9.97 & 16.67 & 0.598 & $-38.3$ & 0.079 & -- \\
HAR + Statistics & 9.85 & 17.00 & 0.579 & $-37.9$ & 0.071 & 0.44 \\
HAR + log-sig. & 9.81 & 17.24 & 0.569 & $-38.3$ & 0.066 & 0.28 \\
HAR + raw sig. & 9.39 & 17.08 & 0.550 & $-42.7$ & 0.058 & 0.24 \\
\addlinespace
\multicolumn{7}{@{}l}{\textit{Panel B. Developed markets excluding the United States}} \\
HAR & 5.18 & 15.30 & 0.339 & $-49.1$ & 0.068 & -- \\
HAR + Statistics & 5.02 & 15.47 & 0.324 & $-48.8$ & 0.061 & 0.59 \\
HAR + log-sig. & 6.08 & 15.52 & 0.392 & $-48.7$ & 0.059 & 0.02 \\
HAR + raw sig. & 5.65 & 16.04 & 0.352 & $-55.5$ & 0.056 & 0.78 \\
\bottomrule
\end{tabular*}
\notes{The exposure rule of Table~\ref{tab:econ} driven by the tercile probabilities of the HAR regressions of Table~\ref{tab:harsig}, with the scale $c$ set on the training window. The $p$-value tests equality of the rule's Sharpe ratio and that of the plain HAR rule using the method of \citet{ledoit2008}; none is resolved after Holm correction across the six comparisons. This evaluation was added after the forecasts of Table~\ref{tab:harsig} had been scored.}
\end{table}

\section{When do signatures help? Controlled experiments}\label{sec:mechanism}
The forecasting results leave open why signature gains are modest and baseline-dependent. We answer with two controlled Heston designs in which the source of regime differences is known: in Design A regimes differ only in variance scale, and in Design B they share one stationary variance law and differ only in speed.

\subsection{Two Heston regime designs}\label{sec:designs}
For an exogenous regime process $Z_t\in\{0,1,2\}$, log prices follow the stochastic-volatility model of \citet{heston1993}:
\begin{align}
 \mathrm{d}V_t&=\kappa_{Z_t}(\theta_{Z_t}-V_t)\,\mathrm{d}t
 +\sigma_{Z_t}\sqrt{V_t}\,\mathrm{d}W_t^V,\nonumber\\
 \mathrm{d}\log S_t&=(\mu_{Z_t}-V_t/2)\,\mathrm{d}t
 +\sqrt{V_t}\,\mathrm{d}W_t^S,\qquad
 \mathrm{d}\langle W^S,W^V\rangle_t=\rho_{Z_t}\,\mathrm{d}t.
 \label{eq:heston}
\end{align}
Each simulated path has 5,000 daily observations with $\Delta=1/252$. Regime changes are evaluated once per day. Each state persists for at least 50 observations; thereafter it switches with probability 0.002 per observation, with the new state drawn uniformly from the other two states. Parameters are drawn per path within the design constraints (Supporting Information, Table S1).

We use an Euler--Maruyama scheme with absorption at zero \citep{lord2010}: the variance is updated with coefficients evaluated at $V_t^+=\max(V_t,0)$ and then projected onto the non-negative half-line. Design A uses one step per day, whereas Design B uses four equal substeps per day to reduce discretization error under its fastest mean-reversion rates. Initial variance is drawn from the invariant CIR distribution of the initial state and is not reset following regime changes.

In \emph{Design A} (scale), $\mu,\kappa,\rho,c$ are common across states, $\theta_z$ varies and $\sigma_z=c\sqrt{\kappa\theta_z}$. Within a state, $U_t=V_t/\theta_z$ satisfies $\dd U_t=\kappa(1-U_t)\dd t+c\sqrt{\kappa U_t}\dd W_t^V$, so normalized dynamics are identical and states differ only in variance scale.

In \emph{Design B} (speed), $\theta,\mu,\rho,q$ are common, $\sigma_z=\sqrt{q\kappa_z}$, and $\kappa_z$ lies in $[0.8,1.2]$, $[4.5,6.0]$ or $[14,18]$. Every state then has the same stationary variance law, and generators differ only by the factor $\kappa_z$:
\begin{equation}
 \pi=\operatorname{Gamma}(2\theta/q,\,q/2),\qquad
 \mathcal L_z f(v)=\kappa_z\left[(\theta-v)f'(v)+\frac{qv}{2}f''(v)\right],
 \label{eq:control}
\end{equation}
with Gamma shape and scale parameters. Under independent switching and stationary initialization, every state-specific transition kernel preserves $\pi$. The implied variance half-lives, $\ln 2/\kappa_z$, are about 146--218, 29--39 and 10--12 trading days, so a 50-observation window spans only a fraction of the slow state's memory.

In both designs the Feller condition $2\kappa\theta\geq\sigma^2$ is state-independent: it reduces to $c^2\leq 2$ in Design A; because \(c\in[0.45,0.55]\), the Feller ratio \(2/c^2\) lies between 6.61 and 9.88. In Design B it reduces to \(q\leq2\theta\); with \(\theta\in[0.07,0.09]\) and \(q\in[0.035,0.055]\), the stationary Gamma shape \(2\theta/q\) is at least 2.54. The control concerns instantaneous variance, not every return distribution: finite-horizon return moments depend on temporal covariance and can differ across states (Supporting Information, Section S1). Discretization also breaks the matching slightly: with four Euler substeps, the stationary variance of $V$ differs across the three Design B states by about one percentage point, against about three points with a single step (Supporting Information, Section S1).

\subsection{From nowcasts to forecasts}\label{sec:nowcast}
Each simulated window contains the most recent $W=50$ returns, is represented by the predictor sets of Section~\ref{sec:predictors} without the HAR averages, and is labeled by the regime at its endpoint. The simulated task is therefore a nowcast: identifying the current regime from past returns alone. Because regimes are persistent, the regime one month ahead equals the current regime for more than 95\% of windows, and a short-horizon regime forecast is essentially the nowcast propagated through the switching law. The quality of a representation's nowcast therefore largely determines the quality of its short-horizon forecast, and the two designs isolate which kind of regime difference, in level or in dynamics, a representation can identify.

The same three learners and hyperparameter grids as in Section~\ref{sec:evaluation} are used, with the setting of highest validation balanced accuracy selected separately for every predictor set (Supporting Information, Section S2). The hidden Markov benchmark is fitted to the training return sequences with five EM initializations, and its hidden states are matched to regime labels using training windows only. We report both its direct classification, from the largest filtered probability, and, for a like-for-like comparison, the accuracy obtained when its three filtered probabilities are supplied to a supervised learner. As an order diagnostic, we also permute the returns within each window before recomputing the signature, which preserves the observations in the window while destroying their ordering.

Each of 50 independent replications contains 10 training, 10 validation and 5 test paths. Complete paths are assigned to these sets before window construction, producing 49,510 training windows, 49,510 validation windows and 24,755 test windows per replication; windows overlap only within a path, and neither latent variance nor the data-generating parameters are used as predictors. For an augmentation, the paired contrast in replication \(i\) is
\[
D_i=
\operatorname{BA}_{i,\mathrm{augmented}}
-\operatorname{BA}_{i,\mathrm{base}}.
\]
We report its mean and the Student-\(t\) interval
\[
\bar D
\pm
t_{49,0.975}\frac{s_D}{\sqrt{50}},
\]
together with the number of replications for which \(D_i>0\). These intervals summarize variation across the complete simulation-and-fitting procedure; they are not estimates of forecast uncertainty in deployment. Holm-adjusted $p$-values for every reported contrast are given in the Supporting Information, Section S3.

\subsection{Results}\label{sec:simresults}
Table~\ref{tab:grid} reports balanced accuracy for every representation and learner. In Design A, conventional features already separate states well: with the random forest, Statistics reaches 73.49\%, and log-signature augmentation adds 0.47 percentage points [0.39, 0.56], exceeding Statistics in 49 of 50 replications. The direct hidden Markov filter reaches 72.33\%, below the supervised conventional-feature models. Against B3, the largest signature gain is 1.04 points [0.87, 1.20], obtained by adding raw signatures to logistic regression. Thus, signatures provide modest but systematic increments when regimes differ principally in variance scale.

Design B is substantially harder. With the random forest, Statistics reaches 39.48\% against a chance level of 33.33\%, and log-signature augmentation adds 1.16 points [0.88, 1.44], positive in 46 of 50 replications. The direct hidden Markov filter reaches only 34.43\%, consistent with a Gaussian-emission model being unable to represent differences carried mainly by variance dynamics rather than variance level.

The signature representation interacts strongly with the learner. In Design B, adding raw signatures to B3 increases logistic-regression accuracy by 4.06 points [3.70, 4.42], compared with 0.41 points [0.22, 0.60] for the random forest and 0.53 points [0.37, 0.69] for gradient boosting. The corresponding log-signature gains are 1.45, 0.35 and 0.48 points. This pattern is consistent with linear functionals of raw-signature coordinates being directly accessible to logistic regression, whereas the transformation between raw and log-signature coordinates is nonlinear.

Comparing the two designs isolates what the signature contributes. The augmentation gain is larger in the speed design than in the scale design in eleven of the twelve augmentation contrasts, and the exception, raw signatures added to B3 under the random forest, is small in both. Signatures therefore add information about how variance evolves within the window rather than about its level. This is consistent with Section~\ref{sec:har}, where signatures improve the HAR regression but conventional features do not: the HAR regression already captures the level of volatility at several horizons, and signatures add information about its dynamics.

The same comparison reveals a small inversion in the conventional baselines. Although B3 contains Statistics, it performs slightly worse under logistic regression in both designs: 72.61\% versus 73.03\% in Design A and 39.29\% versus 39.78\% in Design B. Expanding the logistic penalty search from three to nine values leaves the inversion essentially unchanged (Supporting Information, Section S4), which suggests that the additional conventional predictors add estimation noise for the linear learner rather than that the regularization grid is too narrow. The inversion is minor in the simulation, where every replication supplies independent training paths, but it is large in the United States forecasts of Section~\ref{sec:forecasts}, where it governs how the market contrasts should be read.

The order diagnostic confirms that temporal ordering contributes to these gains. In Design B, replacing the ordered raw signature with a within-window shuffled raw signature reduces B3-augmented accuracy by 2.50 points [2.19, 2.81] for logistic regression and 0.32 points [0.14, 0.51] for the random forest. However, shuffled raw signatures still add 1.56 points [1.32, 1.80] over B3 for logistic regression, so order explains a substantial, but not complete, part of the raw-signature gain (Supporting Information, Table S6). The truncation level also matters differently across designs: lower levels capture most of the gain in Design A, whereas higher levels add information in Design B (Supporting Information, Section S5).

\begin{table}[htbp]
\centering
\caption{Simulated state classification by representation, learner and
regime design.}
\label{tab:grid}
\scriptsize
\begin{tabular*}{\linewidth}
{@{\extracolsep{\fill}}lcccccc@{}}
\toprule
& \multicolumn{3}{c}{Design A: scale}
& \multicolumn{3}{c}{Design B: speed}\\
\cmidrule(lr){2-4}\cmidrule(l){5-7}
Representation
& Logit & RF & GBM
& Logit & RF & GBM\\
\midrule

\multicolumn{7}{@{}l}
{\emph{Panel A. Balanced accuracy (\%)}}\\

HMM direct
& \multicolumn{3}{c}{72.33}
& \multicolumn{3}{c}{34.43}\\

HMM probability block
& 72.56 & 72.64 & 72.63
& 34.80 & 35.49 & 35.42\\

Statistics
& 73.03 & 73.49 & 72.82
& 39.78 & 39.48 & 39.28\\

Statistics + log-sig.
& 73.85 & 73.96 & 73.56
& 41.66 & 40.63 & 40.72\\

Statistics + raw sig.
& 74.33 & 74.35 & 73.66
& 44.31 & 40.76 & 40.74\\

B3
& 72.61 & 73.78 & 72.84
& 39.29 & 40.41 & 39.87\\

B3 + log-sig.
& 73.10 & 73.97 & 73.26
& 40.74 & 40.76 & 40.34\\

B3 + raw sig.
& 73.65 & 74.34 & 73.33
& 43.35 & 40.82 & 40.40\\

\midrule

\multicolumn{7}{@{}l}
{\emph{Panel B. Paired gain over the corresponding base block:
percentage points [95\% CI]; positive replications}}\\

Statistics $\to$ + log-sig.
& \shortstack{\(+0.83\)\\ \([0.68,0.97]\)\\ \(46/50\)}
& \shortstack{\(+0.47\)\\ \([0.39,0.56]\)\\ \(49/50\)}
& \shortstack{\(+0.74\)\\ \([0.63,0.86]\)\\ \(49/50\)}
& \shortstack{\(+1.88\)\\ \([1.59,2.17]\)\\ \(49/50\)}
& \shortstack{\(+1.16\)\\ \([0.88,1.44]\)\\ \(46/50\)}
& \shortstack{\(+1.44\)\\ \([1.20,1.68]\)\\ \(49/50\)}\\

Statistics $\to$ + raw sig.
& \shortstack{\(+1.30\)\\ \([1.09,1.52]\)\\ \(49/50\)}
& \shortstack{\(+0.86\)\\ \([0.73,0.99]\)\\ \(50/50\)}
& \shortstack{\(+0.85\)\\ \([0.71,0.99]\)\\ \(49/50\)}
& \shortstack{\(+4.53\)\\ \([4.07,4.99]\)\\ \(50/50\)}
& \shortstack{\(+1.28\)\\ \([0.94,1.62]\)\\ \(44/50\)}
& \shortstack{\(+1.46\)\\ \([1.20,1.71]\)\\ \(47/50\)}\\

B3 $\to$ + log-sig.
& \shortstack{\(+0.49\)\\ \([0.37,0.61]\)\\ \(43/50\)}
& \shortstack{\(+0.19\)\\ \([0.11,0.27]\)\\ \(37/50\)}
& \shortstack{\(+0.42\)\\ \([0.34,0.49]\)\\ \(48/50\)}
& \shortstack{\(+1.45\)\\ \([1.23,1.67]\)\\ \(48/50\)}
& \shortstack{\(+0.35\)\\ \([0.18,0.53]\)\\ \(34/50\)}
& \shortstack{\(+0.48\)\\ \([0.33,0.63]\)\\ \(40/50\)}\\

B3 $\to$ + raw sig.
& \shortstack{\(+1.04\)\\ \([0.87,1.20]\)\\ \(48/50\)}
& \shortstack{\(+0.56\)\\ \([0.46,0.67]\)\\ \(48/50\)}
& \shortstack{\(+0.49\)\\ \([0.40,0.57]\)\\ \(47/50\)}
& \shortstack{\(+4.06\)\\ \([3.70,4.42]\)\\ \(49/50\)}
& \shortstack{\(+0.41\)\\ \([0.22,0.60]\)\\ \(35/50\)}
& \shortstack{\(+0.53\)\\ \([0.37,0.69]\)\\ \(39/50\)}\\

\bottomrule
\end{tabular*}

\notes{Panel A reports mean test balanced accuracy across 50 independent paired replications; chance accuracy is 33.33\%. Panel B reports the mean within-replication difference between an augmented block and its corresponding conventional baseline, its nominal 95\% Student-\(t\) interval, and the number of replications with a positive difference. HMM direct classifies each window using the largest forward-filtered state probability. The HMM probability block supplies the three filtered probabilities to the indicated supervised learner. Logit: \(\ell_2\)-regularized logistic regression; RF: random forest; GBM: histogram gradient boosting. Each learner uses a prespecified three-setting hyperparameter grid selected on validation balanced accuracy. B3 contains all 11 Statistics features plus 29 additional conventional features.}
\end{table}

\subsection{Timeliness of regime signals}\label{sec:timeliness}
A forecaster who acts on regime changes also needs timely signals. In \ref{app:detection}, we convert each classifier's probabilities into a causal change detector and compare detectors at matched false-alarm rates against a cumulative-sum benchmark \citep{page1954}. Signature augmentation does not make detection timelier: at one false alarm per 1,000 observations no augmentation yields a nominally resolved gain in detection probability, only five of 32 detection contrasts are nominally resolved, all for raw signatures at two or five false alarms per 1,000, and none survives Holm correction. Consistent with this, the classification gains arise mainly in established regimes: for log-signatures with a fixed-depth random forest, no gain is resolved in the first 25 observations after a switch (Supporting Information, Table S10). Signatures therefore help to identify, and hence to forecast, persistent regimes rather than to signal their onset.

\section{Conclusion}\label{sec:conclusion}
Path signatures improve volatility-regime forecasts under specific and checkable conditions. In daily equity data, machine-learning classifiers, with or without signatures, do not beat a HAR regression of realized volatility, but adding raw signatures to the HAR regression itself improves United States forecasts of next-month volatility terciles in balanced accuracy and in the ranked probability score, whereas adding conventional volatility features does not; outside the United States no gain survives correction for multiplicity. Controlled Heston experiments explain why signatures, rather than conventional features, carry this information: signatures describe how regimes evolve rather than how volatile they are, so they add modestly when regimes differ in variance level, which realized-volatility predictors already capture, and more when regimes differ in mean-reversion speed, most of all when raw signatures feed a linear model. Better identification of established regimes does not, however, produce earlier signals of regime change, and no accuracy gain translates into a resolved Sharpe-ratio improvement in a volatility-managed portfolio.

For forecasting practice, these results suggest adding signature coordinates to a strong realized-volatility benchmark rather than replacing the benchmark with a classifier, evaluating them against compact as well as rich conventional baselines, choosing the signature coordinate system together with the model, and judging any accuracy gain by the decision it is meant to inform, since gains of the size found here need not change portfolio outcomes. Studies that use regime forecasts to time decisions should evaluate the onset of regimes separately, at explicit false-alarm budgets and with multiplicity-aware inference. Our conclusions are conditional on the simulated parameter families, numerical discretization, window length, learners, forecast horizon and market samples considered; the signature-augmented HAR comparison was designed after the benchmark results were known, and its economic evaluation was added after the forecasts had been scored.

\FloatBarrier
\section*{Acknowledgements}
The authors used ChatGPT (OpenAI) and Claude (Anthropic) to restructure drafts, edit language, check references and prepare the LaTeX source, and Claude to assist with implementing portions of the analysis code. The authors designed the study, reviewed and tested all code, verified every reported result against the archived replication-level outputs, and take full responsibility for the content of the article.

\section*{Author contributions}
\textbf{Yang Lyu:} Conceptualization, Methodology, Software, Formal analysis, Investigation, Data curation, Visualization, Writing -- original draft. \textbf{Kiseop Lee:} Conceptualization, Methodology, Supervision, Writing -- review and editing.

\section*{Conflict of interest statement}
The authors declare that they have no known competing financial interests or personal relationships that could have appeared to influence the work reported in this paper.

\section*{Funding}
This research did not receive any specific grant from funding agencies in the public, commercial, or not-for-profit sectors.

\section*{Data availability statement}
The code, configuration files and replication-level outputs supporting the forecasting, simulation and detection analyses are available at \url{https://github.com/letdaly/regime_signature}; the per-replication probability and path archives used by the detection study are attached to release v1.0 of the repository. The market return series are publicly available from the Kenneth French Data Library \citep{french2026}; the repository records the sample definition and SHA-256 hashes of the source files used.

\bibliographystyle{elsarticle-harv}
\bibliography{references}

\appendix
\setcounter{figure}{0}
\renewcommand{\thefigure}{A.\arabic{figure}}
\renewcommand{\theHfigure}{A.\arabic{figure}}
\section{Regime-change detection at matched false-alarm rates}\label{app:detection}
\subsection{Detection protocol}
We convert the probability sequence from each classifier into a causal regime-change detector. Predicted probabilities are exponentially smoothed, and the detector declares a switch when a state different from the current state has the largest smoothed probability, exceeds a confidence threshold, and remains the candidate state for a prescribed number of consecutive observations. The hidden Markov filtered probabilities are passed through the same detection rule.

As a classical benchmark, a two-sided CUSUM detector \citep{page1954} monitors log squared returns relative to an exponentially weighted reference level. Its allowance, reference half-life and decision limit are selected on validation paths. The CUSUM identifies upward or downward changes in return variation but does not assign one of the three regime labels.

For each representation and learner, the smoothing rate, confidence threshold and confirmation count are calibrated on validation paths. We consider target false-alarm budgets of 0.5, 1, 2 and 5 false alarms per 1,000 scored observations. At each target, we select the setting with the highest validation detection probability among those whose validation false-alarm rate does not exceed the target, and apply it once to the test paths; the same validation-only rule chooses the CUSUM parameters. Because the targets are imposed on validation paths, realized test false-alarm rates need not equal them exactly and are reported alongside all test detection results (Supporting Information, Section S6).

An alarm is matched one-to-one to the earliest unmatched true change that precedes it by no more than 100 observations. The matching is state-agnostic, so that CUSUM and probability-based detectors are evaluated on the same event-timing objective. Unmatched alarms are counted as false alarms. Detection probability, conditional detection delay and false alarms per 1,000 observations are computed over the same test-window endpoints, beginning after the common 50-observation warm-up period.

\subsection{Results}
\begin{figure}[htbp]
\centering
\includegraphics[width=\linewidth]{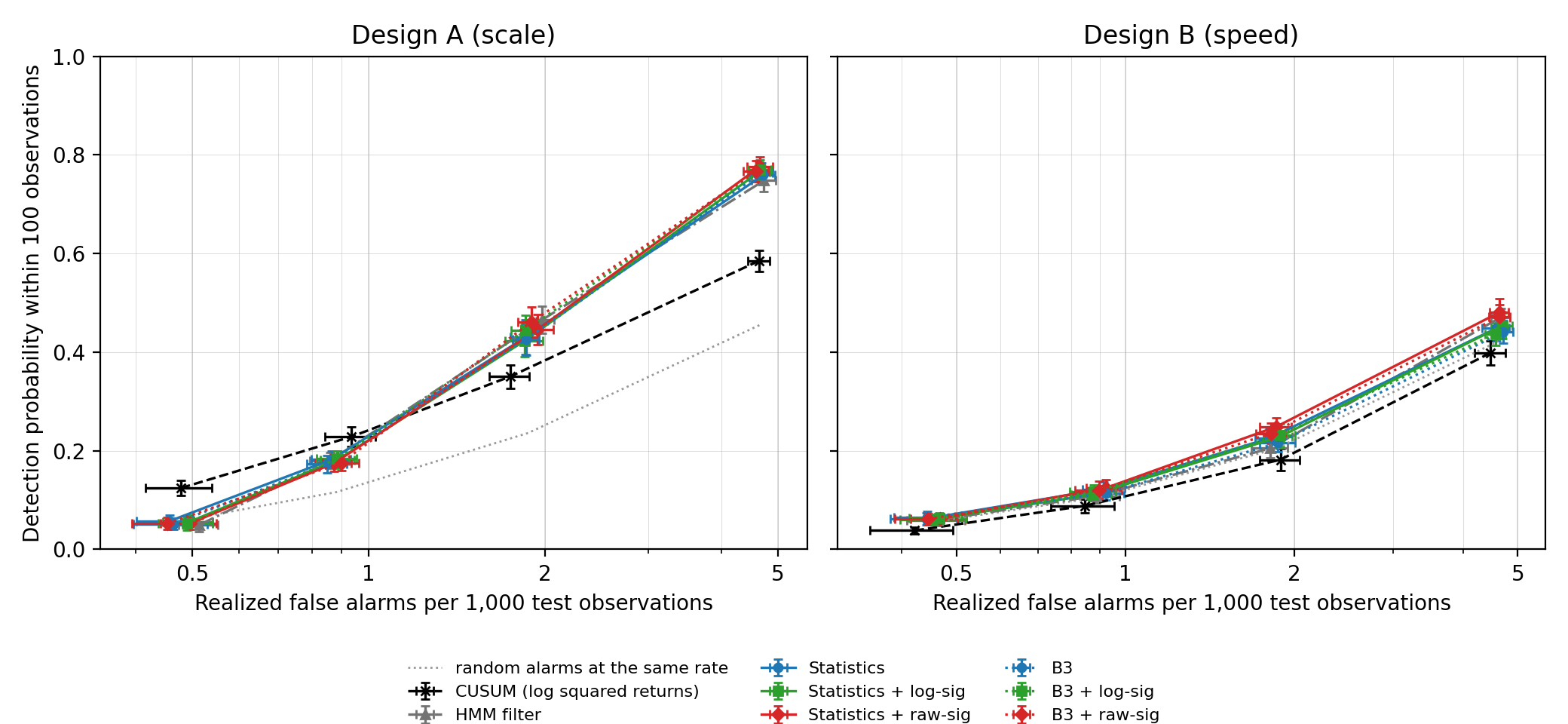}
\caption{Regime-change detection under common false-alarm budgets. Detection probability within 100 observations of a true change is plotted against the realized number of false alarms per 1,000 test observations for Design A (left) and Design B (right). Detector settings are selected on validation paths to target 0.5, 1, 2 and 5 false alarms per 1,000 observations and are then applied once to test paths. Because calibration is performed on validation paths, realized test rates need not equal their targets exactly. Points are means over 50 paired replications; horizontal and vertical bars are nominal 95\% replication-level intervals. The dotted curve gives the detection probability of a memoryless random-alarm process at the corresponding alarm rate. CUSUM is a two-sided cumulative-sum detector applied to log squared returns; HMM denotes the forward-filtered Gaussian hidden Markov model. For each conventional or signature representation, the classifier learner is selected by validation balanced accuracy.}
\label{fig:tradeoff}
\end{figure}

Figure~\ref{fig:tradeoff} plots the probability of detecting a regime change within 100 observations against the realized test false-alarm rate. The ranking of the detectors depends on the false-alarm budget. In Design A, at the target of one false alarm per 1,000 observations, CUSUM detects 22.8\% [20.8, 24.9] of changes, compared with 18.0\% [16.3, 19.6] for the hidden Markov filter, 18.2\% [16.5, 19.9] for Statistics and 18.2\% [16.4, 19.9] for Statistics plus log-signatures. Among the detectors shown, CUSUM performs best at this tight budget, whereas the classifier-based and hidden Markov detectors overtake it at targets of two and five false alarms per 1,000 observations.

Design B is more difficult at the event level. At the target of one false alarm per 1,000 observations, the best representation, Statistics plus raw signatures, detects 12.4\% [10.8, 14.1] of changes. Statistics detects 12.2\% [10.7, 13.7], the hidden Markov filter 10.8\% [9.4, 12.3], and CUSUM 8.8\% [7.4, 10.2]. CUSUM performs at approximately the memoryless-alarm reference because the regimes share a stationary variance law and differ primarily through temporal dynamics.

At the target of one false alarm per 1,000 observations, neither raw nor log-signature augmentation produces a nominally resolved increase in detection probability for the validation-selected learner in either design. At the target of five, three raw-signature contrasts are nominally resolved: Statistics plus raw signatures adds 1.77 percentage points [0.40, 3.14] in Design A and 3.24 points [0.05, 6.43] in Design B, while B3 plus raw signatures adds 3.17 points [0.37, 5.97] in Design B. Across all designs, budgets and augmentations, however, only five of 32 detection contrasts are nominally resolved and none survives Holm correction. Nor does any selected-learner comparison yield a resolved reduction in conditional detection delay.

A time-since-change diagnostic for the fixed-depth random-forest log-signature specification helps explain this gap. No gain is resolved during the first 25 observations after a switch; the clearest gains occur at ages of 100 observations or more in Design A, at 0.34 points [0.27, 0.42], and at ages 50--99 in Design B, at 1.31 points [0.83, 1.78] (Supporting Information, Table S10). In Design A, the label changes immediately but variance relaxes toward the new $\theta_z$ at rate $\kappa$. Conditional mean delays of about 64 observations at the one-per-1,000 target are therefore compatible with both this transition and the detector's smoothing and confirmation rules.

\end{document}


\maketitle

\noindent Tables S3--S14 are generated from the replication-level outputs by
\path{paper/supplement/build_supplement_tables.py}, and Figure~1 of the main text by
\path{paper/supplement/build_b3_figure.py}; no model is refitted. Unless
stated otherwise, intervals are nominal 95\% Student-$t$ intervals over 50 paired
replications, differences are in percentage points, and bold marks an interval that
excludes zero.

\section{Simulation design}\label{sec:s-design}

\subsection{Parameter families}

Each path receives a fresh parameter draw from the families in Table~\ref{tab:s-params}.
Parameters shared across states are drawn once per path; state-specific parameters are
drawn independently for each state. All draws are uniform on the stated intervals.

\begin{table}[htbp]
\centering
\caption{Parameter families of the two Heston regime designs.}
\label{tab:s-params}
\small
\begin{tabular}{@{}lll@{}}
\toprule
 & Design A (scale) & Design B (speed) \\
\midrule
\multicolumn{3}{@{}l}{\textit{Common to all states within a path}} \\
Drift $\mu$ & $[0.01, 0.04]$ & $[0.01, 0.04]$ \\
Correlation $\rho$ & $[-0.80, -0.65]$ & $[-0.80, -0.65]$ \\
Mean reversion $\kappa$ & $[3.5, 4.5]$ & -- \\
Scale coefficient $c$ & $[0.45, 0.55]$ & -- \\
Long-run variance $\theta$ & -- & $[0.07, 0.09]$ \\
Variance ratio $q=\sigma_z^2/\kappa_z$ & -- & $[0.035, 0.055]$ \\
\addlinespace
\multicolumn{3}{@{}l}{\textit{State-specific}} \\
State 0 & $\theta_0\in[0.015, 0.025]$ & $\kappa_0\in[0.8, 1.2]$ \\
State 1 & $\theta_1\in[0.065, 0.095]$ & $\kappa_1\in[4.5, 6.0]$ \\
State 2 & $\theta_2\in[0.14, 0.20]$ & $\kappa_2\in[14, 18]$ \\
Volatility of variance & $\sigma_z=c\sqrt{\kappa\theta_z}$ & $\sigma_z=\sqrt{q\kappa_z}$ \\
\addlinespace
\multicolumn{3}{@{}l}{\textit{Simulation}} \\
Observations per path & 5,000 & 5,000 \\
Step $\Delta$ & $1/252$ & $1/252$ \\
Euler substeps per observation & 1 & 4 \\
Minimum dwell; switching probability & 50; 0.002 & 50; 0.002 \\
Initial variance & \multicolumn{2}{l}{Invariant CIR law of the initial state} \\
\bottomrule
\end{tabular}
\end{table}

\subsection{What the Design B control does and does not match}

Within state $z$ of Design B, the variance is a stationary CIR process with invariant law
$\pi=\operatorname{Gamma}(2\theta/q,\,q/2)$, mean $\theta$ and variance $\theta q/2$, and
autocovariance $(\theta q/2)\,e^{-\kappa_z|u|}$. The one-time law of $V_t$ is therefore
identical across states. Temporal functionals are not. For the integrated variance
$\mathrm{IV}_h=\int_0^h V_s\,\mathrm{d}s$ over a horizon $h$,
\begin{equation}
 \mathbb{E}[\mathrm{IV}_h]=\theta h,\qquad
 \operatorname{Var}(\mathrm{IV}_h)=\theta q\,\frac{\kappa_z h-1+e^{-\kappa_z h}}{\kappa_z^{2}},
 \label{eq:s-iv}
\end{equation}
so the mean is common but the dispersion falls with $\kappa_z$. Ignoring drift and
leverage ($\rho=0$), the $h$-period log return is conditionally Gaussian given
$\mathrm{IV}_h$, and its excess kurtosis is
\begin{equation}
 3\,\frac{\operatorname{Var}(\mathrm{IV}_h)}{(\theta h)^2}
 =\frac{3q}{\theta}\,\frac{\kappa_z h-1+e^{-\kappa_z h}}{(\kappa_z h)^{2}}.
 \label{eq:s-kurt}
\end{equation}
The ratio in \eqref{eq:s-kurt} decreases from $1/2$ as $\kappa_z h$ grows, so over a
50-day window ($h\approx0.2$) the slow state produces markedly fatter-tailed window returns
than the fast state, whereas at the daily horizon the difference is small. Return moments
at finite horizons can therefore differ across states even though the instantaneous
variance law is matched. At representative parameters ($\theta=0.08$, $q=0.045$ and
$\kappa_z=1$, 5.25 and 16, the midpoints of the Design B ranges), the excess kurtosis in
\eqref{eq:s-kurt} is 0.84, 0.84 and 0.83 at the daily horizon but 0.79, 0.61 and 0.37
over a 50-day window. With leverage ($\rho\neq0$) the return distribution is also skewed;
the case $\rho=0$ suffices to show that finite-horizon moments differ across states.

\subsection{Discretization audit}

The Euler scheme with absorption at zero leaves the mean of the variance process
unchanged but, while absorption is inactive, inflates its stationary variance by the factor
$1/(1-\kappa_z\delta/2)$ for a step $\delta$, so faster states drift further from the matched
law. We audited the effect at the midpoints of the Design B ranges ($\theta=0.08$, $q=0.045$
and $\kappa_z=1$, 5.25 and 16). For each state and step size, 100,000 independent variance
paths start from the exact stationary law; the first 10 years of daily observations are
discarded, and the stationary moments are averaged over the next 10 years. With one Euler
step per observation, the stationary variance of $V$ exceeds its exact value $\theta q/2$ by
0.3\%, 0.9\% and 3.2\% in the three states, a spread of about 3.0 percentage points; with four
substeps the deviations are $-0.1\%$, 0.0\% and 0.8\%, a spread of about 1.0 point (Monte
Carlo standard errors between 0.05 and 0.2 points). These values agree with the factor above,
the stationary mean stays within 0.1\% of $\theta$, and absorption at zero occurs in fewer than
one step in 10,000. Design B therefore uses four substeps. In Design A all states share
$\kappa$, so the discretized normalized process $V_t/\theta_z$ follows the same recursion in
every state and the scale-only control holds exactly; Design A uses one step. The audit is
reproduced by \path{audit_discretization.py} in the replication repository.

\clearpage
\section{Model and detector specifications}\label{sec:s-models}

Table~\ref{tab:s-specs} lists every tuned setting. Each learner searches a three-point
grid, and the setting with the highest validation balanced accuracy is selected
separately for each feature block and replication. All learners are class-balanced.

\begin{table}[htbp]
\centering
\caption{Learner, benchmark and detector specifications.}
\label{tab:s-specs}
\small
\begin{tabular}{@{}lp{0.62\textwidth}@{}}
\toprule
Component & Specification \\
\midrule
Logistic regression & $\ell_2$ penalty, inverse strength $C\in\{0.01, 0.1, 1\}$; features
scaled robustly (10th--90th percentile range) and clipped at $\pm20$ \\
Random forest & 200 trees, maximum depth $\in\{4, 6, 10\}$ \\
Gradient boosting & Histogram gradient boosting, maximum leaf nodes $\in\{7, 15, 31\}$,
learning rate 0.05, 300 iterations \\
Gaussian HMM & Three states, returns in percent, EM with at most 200 iterations and five
initializations (one persistent, diagonal 0.97), fit with the highest training
log likelihood retained; states matched to labels on training windows by the Hungarian
algorithm \\
\addlinespace
Probability detector & Exponential smoothing rate $\in\{0.2, 0.5, 1\}$; confirmation count
$\in\{1, 3, 5\}$; 24 confidence thresholds from 0.34 to 0.99 \\
CUSUM & Two-sided, on log squared returns relative to an exponentially weighted reference;
allowance $\in\{0.25, 0.5, 1\}$; reference half-life $\in\{50, 250\}$ observations; 24
geometrically spaced decision limits from 2 to 200; sums reset and the reference
re-anchors after an alarm \\
Event matching & One-to-one, earliest unmatched true change at most 100 observations
before the alarm \\
\bottomrule
\end{tabular}
\end{table}

The forecasting study (Section~2 of the main text) uses the same learners and grids; the
setting and the learner are selected on the last three years of each training window.

\clearpage
\section{Multiple-testing correction}\label{sec:s-holm}

Tables~\ref{tab:s-holm-class} and~\ref{tab:s-holm-det} report 64 contrasts from
simulated classification (24), market classifiers (eight) and regime-change detection
(32). They give unadjusted $p$-values and Holm-adjusted $p$-values within each family
and across these 64 contrasts. Simulation $p$-values are from paired $t$-tests over
50 replications; market-classifier balanced-accuracy $p$-values are from the stationary
bootstrap (2,000 resamples, mean block length 21 days). The HAR-augmentation tests
form separate six-comparison families for each score and are reported in
Table~\ref{tab:s-har-pvals}.

\begin{table}[htbp]
\centering
\caption{Classification and market contrasts with Holm-adjusted $p$-values.}
\label{tab:s-holm-class}
\resizebox{\textwidth}{!}{\begin{tabular}{@{}lllcccc@{}}
\toprule
Design / market & Learner & Contrast & Gain [95\% CI] & $p$ & $p_{\mathrm{Holm}}^{\mathrm{family}}$ & $p_{\mathrm{Holm}}^{\mathrm{all}}$ \\
\midrule
\multicolumn{7}{@{}l}{\textit{Panel A. Simulated classification (family of 24)}} \\
Design A & Logit & Statistics $\to$ + log-sig. & \textbf{\boldmath +0.83 [0.68, 0.97]} & $<$0.001 & $<$0.001 & $<$0.001 \\
Design A & Logit & Statistics $\to$ + raw sig. & \textbf{\boldmath +1.30 [1.09, 1.52]} & $<$0.001 & $<$0.001 & $<$0.001 \\
Design A & Logit & B3 $\to$ + log-sig. & \textbf{\boldmath +0.49 [0.37, 0.61]} & $<$0.001 & $<$0.001 & $<$0.001 \\
Design A & Logit & B3 $\to$ + raw sig. & \textbf{\boldmath +1.04 [0.87, 1.20]} & $<$0.001 & $<$0.001 & $<$0.001 \\
Design A & RF & Statistics $\to$ + log-sig. & \textbf{\boldmath +0.47 [0.39, 0.56]} & $<$0.001 & $<$0.001 & $<$0.001 \\
Design A & RF & Statistics $\to$ + raw sig. & \textbf{\boldmath +0.86 [0.73, 0.99]} & $<$0.001 & $<$0.001 & $<$0.001 \\
Design A & RF & B3 $\to$ + log-sig. & \textbf{\boldmath +0.19 [0.11, 0.27]} & $<$0.001 & $<$0.001 & 0.002 \\
Design A & RF & B3 $\to$ + raw sig. & \textbf{\boldmath +0.56 [0.46, 0.67]} & $<$0.001 & $<$0.001 & $<$0.001 \\
Design A & GBM & Statistics $\to$ + log-sig. & \textbf{\boldmath +0.74 [0.63, 0.86]} & $<$0.001 & $<$0.001 & $<$0.001 \\
Design A & GBM & Statistics $\to$ + raw sig. & \textbf{\boldmath +0.85 [0.71, 0.99]} & $<$0.001 & $<$0.001 & $<$0.001 \\
Design A & GBM & B3 $\to$ + log-sig. & \textbf{\boldmath +0.42 [0.34, 0.49]} & $<$0.001 & $<$0.001 & $<$0.001 \\
Design A & GBM & B3 $\to$ + raw sig. & \textbf{\boldmath +0.49 [0.40, 0.57]} & $<$0.001 & $<$0.001 & $<$0.001 \\
Design B & Logit & Statistics $\to$ + log-sig. & \textbf{\boldmath +1.88 [1.59, 2.17]} & $<$0.001 & $<$0.001 & $<$0.001 \\
Design B & Logit & Statistics $\to$ + raw sig. & \textbf{\boldmath +4.53 [4.07, 4.99]} & $<$0.001 & $<$0.001 & $<$0.001 \\
Design B & Logit & B3 $\to$ + log-sig. & \textbf{\boldmath +1.45 [1.23, 1.67]} & $<$0.001 & $<$0.001 & $<$0.001 \\
Design B & Logit & B3 $\to$ + raw sig. & \textbf{\boldmath +4.06 [3.70, 4.42]} & $<$0.001 & $<$0.001 & $<$0.001 \\
Design B & RF & Statistics $\to$ + log-sig. & \textbf{\boldmath +1.16 [0.88, 1.44]} & $<$0.001 & $<$0.001 & $<$0.001 \\
Design B & RF & Statistics $\to$ + raw sig. & \textbf{\boldmath +1.28 [0.94, 1.62]} & $<$0.001 & $<$0.001 & $<$0.001 \\
Design B & RF & B3 $\to$ + log-sig. & \textbf{\boldmath +0.35 [0.18, 0.53]} & $<$0.001 & $<$0.001 & 0.009 \\
Design B & RF & B3 $\to$ + raw sig. & \textbf{\boldmath +0.41 [0.22, 0.60]} & $<$0.001 & $<$0.001 & 0.004 \\
Design B & GBM & Statistics $\to$ + log-sig. & \textbf{\boldmath +1.44 [1.20, 1.68]} & $<$0.001 & $<$0.001 & $<$0.001 \\
Design B & GBM & Statistics $\to$ + raw sig. & \textbf{\boldmath +1.46 [1.20, 1.71]} & $<$0.001 & $<$0.001 & $<$0.001 \\
Design B & GBM & B3 $\to$ + log-sig. & \textbf{\boldmath +0.48 [0.33, 0.63]} & $<$0.001 & $<$0.001 & $<$0.001 \\
Design B & GBM & B3 $\to$ + raw sig. & \textbf{\boldmath +0.53 [0.37, 0.69]} & $<$0.001 & $<$0.001 & $<$0.001 \\
\addlinespace
\multicolumn{7}{@{}l}{\textit{Panel B. Market forecasts, validation-selected learner (family of 8)}} \\
United States & Selected & Statistics $\to$ + log-sig. & +2.13 [$-0.68$, 4.88] & 0.138 & 0.550 & 1.000 \\
United States & Selected & Statistics $\to$ + raw sig. & \textbf{\boldmath +3.68 [1.36, 6.03]} & 0.003 & 0.021 & 0.114 \\
United States & Selected & B3 $\to$ + log-sig. & \textbf{\boldmath +3.60 [1.30, 5.98]} & 0.002 & 0.012 & 0.060 \\
United States & Selected & B3 $\to$ + raw sig. & \textbf{\boldmath +3.72 [0.94, 6.31]} & 0.007 & 0.045 & 0.277 \\
Developed ex US & Selected & Statistics $\to$ + log-sig. & +1.32 [$-1.19$, 3.84] & 0.287 & 0.862 & 1.000 \\
Developed ex US & Selected & Statistics $\to$ + raw sig. & $-0.49$ [$-3.70$, 2.88] & 0.776 & 0.862 & 1.000 \\
Developed ex US & Selected & B3 $\to$ + log-sig. & $-1.13$ [$-3.33$, 0.99] & 0.316 & 0.862 & 1.000 \\
Developed ex US & Selected & B3 $\to$ + raw sig. & \textbf{\boldmath $-3.19$ [$-5.82$, $-0.53$]} & 0.018 & 0.092 & 0.647 \\
\bottomrule
\end{tabular}
}
\end{table}

\begin{table}[htbp]
\centering
\caption{Detection contrasts with Holm-adjusted $p$-values. $\Delta$ DP is the paired
difference in the probability of detecting a change within 100 observations, in
percentage points, for the validation-selected learner.}
\label{tab:s-holm-det}
\resizebox{\textwidth}{!}{\begin{tabular}{@{}lllcccc@{}}
\toprule
Design & Target & Contrast & $\Delta$ DP [95\% CI] & $p$ & $p_{\mathrm{Holm}}^{\mathrm{family}}$ & $p_{\mathrm{Holm}}^{\mathrm{all}}$ \\
\midrule
Design A & 0.5 & Statistics $\to$ + log-sig. & $-0.19$ [$-1.40$, 1.01] & 0.747 & 1.000 & 1.000 \\
Design A & 0.5 & Statistics $\to$ + raw sig. & $-0.48$ [$-1.64$, 0.68] & 0.412 & 1.000 & 1.000 \\
Design A & 0.5 & B3 $\to$ + log-sig. & +0.12 [$-0.82$, 1.06] & 0.800 & 1.000 & 1.000 \\
Design A & 0.5 & B3 $\to$ + raw sig. & +0.10 [$-0.96$, 1.15] & 0.856 & 1.000 & 1.000 \\
Design A & 1 & Statistics $\to$ + log-sig. & $-0.04$ [$-1.61$, 1.54] & 0.963 & 1.000 & 1.000 \\
Design A & 1 & Statistics $\to$ + raw sig. & $-0.75$ [$-2.24$, 0.74] & 0.318 & 1.000 & 1.000 \\
Design A & 1 & B3 $\to$ + log-sig. & +1.05 [$-0.34$, 2.44] & 0.135 & 1.000 & 1.000 \\
Design A & 1 & B3 $\to$ + raw sig. & +0.23 [$-1.58$, 2.04] & 0.798 & 1.000 & 1.000 \\
Design A & 2 & Statistics $\to$ + log-sig. & $-0.63$ [$-3.28$, 2.01] & 0.633 & 1.000 & 1.000 \\
Design A & 2 & Statistics $\to$ + raw sig. & +1.58 [$-1.34$, 4.49] & 0.283 & 1.000 & 1.000 \\
Design A & 2 & B3 $\to$ + log-sig. & +1.91 [$-0.44$, 4.26] & 0.108 & 1.000 & 1.000 \\
Design A & 2 & B3 $\to$ + raw sig. & \textbf{\boldmath +3.54 [1.35, 5.73]} & 0.002 & 0.066 & 0.081 \\
Design A & 5 & Statistics $\to$ + log-sig. & +0.65 [$-0.72$, 2.02] & 0.343 & 1.000 & 1.000 \\
Design A & 5 & Statistics $\to$ + raw sig. & \textbf{\boldmath +1.77 [0.40, 3.14]} & 0.012 & 0.387 & 0.449 \\
Design A & 5 & B3 $\to$ + log-sig. & +0.62 [$-0.25$, 1.49] & 0.157 & 1.000 & 1.000 \\
Design A & 5 & B3 $\to$ + raw sig. & +0.31 [$-0.83$, 1.45] & 0.587 & 1.000 & 1.000 \\
Design B & 0.5 & Statistics $\to$ + log-sig. & $-0.13$ [$-1.75$, 1.48] & 0.867 & 1.000 & 1.000 \\
Design B & 0.5 & Statistics $\to$ + raw sig. & +0.01 [$-1.52$, 1.55] & 0.989 & 1.000 & 1.000 \\
Design B & 0.5 & B3 $\to$ + log-sig. & $-0.17$ [$-1.45$, 1.11] & 0.792 & 1.000 & 1.000 \\
Design B & 0.5 & B3 $\to$ + raw sig. & $-0.32$ [$-1.98$, 1.35] & 0.704 & 1.000 & 1.000 \\
Design B & 1 & Statistics $\to$ + log-sig. & $-1.13$ [$-2.53$, 0.27] & 0.111 & 1.000 & 1.000 \\
Design B & 1 & Statistics $\to$ + raw sig. & +0.25 [$-1.48$, 1.97] & 0.773 & 1.000 & 1.000 \\
Design B & 1 & B3 $\to$ + log-sig. & +0.26 [$-1.10$, 1.62] & 0.703 & 1.000 & 1.000 \\
Design B & 1 & B3 $\to$ + raw sig. & +0.61 [$-1.38$, 2.61] & 0.540 & 1.000 & 1.000 \\
Design B & 2 & Statistics $\to$ + log-sig. & +0.27 [$-2.41$, 2.95] & 0.839 & 1.000 & 1.000 \\
Design B & 2 & Statistics $\to$ + raw sig. & \textbf{\boldmath +2.33 [0.22, 4.44]} & 0.031 & 0.899 & 1.000 \\
Design B & 2 & B3 $\to$ + log-sig. & +1.63 [$-0.44$, 3.70] & 0.120 & 1.000 & 1.000 \\
Design B & 2 & B3 $\to$ + raw sig. & +1.97 [$-0.46$, 4.39] & 0.109 & 1.000 & 1.000 \\
Design B & 5 & Statistics $\to$ + log-sig. & +0.53 [$-2.44$, 3.49] & 0.723 & 1.000 & 1.000 \\
Design B & 5 & Statistics $\to$ + raw sig. & \textbf{\boldmath +3.24 [0.05, 6.43]} & 0.047 & 1.000 & 1.000 \\
Design B & 5 & B3 $\to$ + log-sig. & $-0.36$ [$-2.52$, 1.81] & 0.743 & 1.000 & 1.000 \\
Design B & 5 & B3 $\to$ + raw sig. & \textbf{\boldmath +3.17 [0.37, 5.97]} & 0.027 & 0.815 & 0.924 \\
\bottomrule
\end{tabular}
}
\end{table}

\clearpage
\section{Logistic penalty robustness}\label{sec:s-penalty}

Table~\ref{tab:s-penalty} repeats the logistic cells on the same paths with the inverse
penalty searched over nine values,
\[
C\in\{10^{-4},\ 10^{-3},\ 0.01,\ 0.03,\ 0.1,\ 0.3,\ 1,\ 10,\ 100\}.
\] The upper rows of each panel give test balanced accuracy (\%) under both grids and
their paired difference; the indented rows give the B3-minus-Statistics inversion under
each grid. The wider search moves no cell by more than 0.13 points, and the inversion is
unchanged.

\begin{table}[htbp]
\centering
\caption{Logistic regression under three- and nine-point penalty grids.}
\label{tab:s-penalty}
\resizebox{\textwidth}{!}{\begin{tabular}{@{}lccc@{}}
\toprule
Representation & Three-point grid & Nine-point grid & Nine minus three [95\% CI] \\
\midrule
\multicolumn{4}{@{}l}{\textit{Design A}} \\
Statistics & 73.03 & 73.01 & $-0.01$ [$-0.04$, 0.01] \\
Statistics + log-sig. & 73.85 & 73.84 & $-0.02$ [$-0.06$, 0.02] \\
Statistics + raw sig. & 74.33 & 74.35 & +0.02 [$-0.02$, 0.06] \\
B3 & 72.61 & 72.57 & $-0.04$ [$-0.15$, 0.06] \\
B3 + log-sig. & 73.10 & 73.09 & $-0.01$ [$-0.12$, 0.10] \\
B3 + raw sig. & 73.65 & 73.70 & +0.06 [$-0.06$, 0.17] \\
\quad B3 $-$ Statistics & \textbf{\boldmath $-0.42$ [$-0.60$, $-0.24$]} & \textbf{\boldmath $-0.45$ [$-0.64$, $-0.25$]} &  \\
\quad B3 + log-sig. $-$ Statistics + log-sig. & \textbf{\boldmath $-0.75$ [$-0.92$, $-0.59$]} & \textbf{\boldmath $-0.74$ [$-0.92$, $-0.57$]} &  \\
\quad B3 + raw sig. $-$ Statistics + raw sig. & \textbf{\boldmath $-0.68$ [$-0.83$, $-0.54$]} & \textbf{\boldmath $-0.65$ [$-0.81$, $-0.49$]} &  \\
\addlinespace
\multicolumn{4}{@{}l}{\textit{Design B}} \\
Statistics & 39.78 & 39.65 & $-0.13$ [$-0.31$, 0.05] \\
Statistics + log-sig. & 41.66 & 41.62 & $-0.04$ [$-0.14$, 0.06] \\
Statistics + raw sig. & 44.31 & 44.32 & +0.01 [$-0.03$, 0.05] \\
B3 & 39.29 & 39.16 & $-0.13$ [$-0.30$, 0.05] \\
B3 + log-sig. & 40.74 & 40.74 & $-0.00$ [$-0.05$, 0.04] \\
B3 + raw sig. & 43.35 & 43.30 & \textbf{\boldmath $-0.05$ [$-0.08$, $-0.01$]} \\
\quad B3 $-$ Statistics & \textbf{\boldmath $-0.49$ [$-0.91$, $-0.06$]} & \textbf{\boldmath $-0.48$ [$-0.88$, $-0.09$]} &  \\
\quad B3 + log-sig. $-$ Statistics + log-sig. & \textbf{\boldmath $-0.92$ [$-1.22$, $-0.61$]} & \textbf{\boldmath $-0.88$ [$-1.21$, $-0.55$]} &  \\
\quad B3 + raw sig. $-$ Statistics + raw sig. & \textbf{\boldmath $-0.96$ [$-1.30$, $-0.62$]} & \textbf{\boldmath $-1.02$ [$-1.35$, $-0.68$]} &  \\
\bottomrule
\end{tabular}
}
\end{table}

\clearpage
\section{Ordering and truncation order}\label{sec:s-order}

\subsection{Order diagnostic}

Table~\ref{tab:s-order} compares each augmented block with the same block whose signature
is computed after permuting the returns within each window. Raw and log-signatures share
the permutation.

\begin{table}[htbp]
\centering
\caption{Order diagnostic: ordered versus within-window shuffled signatures.}
\label{tab:s-order}
\resizebox{\textwidth}{!}{\begin{tabular}{@{}lccc@{}}
\toprule
Contrast & Logit & RF & GBM \\
\midrule
\multicolumn{4}{@{}l}{\textit{Design A}} \\
Statistics + log-sig.: ordered $-$ shuffled & \textbf{\boldmath +0.46 [0.35, 0.57]} & \textbf{\boldmath +0.24 [0.16, 0.32]} & \textbf{\boldmath +0.45 [0.36, 0.53]} \\
Statistics + raw sig.: ordered $-$ shuffled & \textbf{\boldmath +0.41 [0.29, 0.54]} & \textbf{\boldmath +0.52 [0.43, 0.60]} & \textbf{\boldmath +0.52 [0.42, 0.62]} \\
B3 + log-sig.: ordered $-$ shuffled & \textbf{\boldmath +0.51 [0.39, 0.63]} & \textbf{\boldmath +0.13 [0.07, 0.20]} & \textbf{\boldmath +0.41 [0.34, 0.49]} \\
B3 + raw sig.: ordered $-$ shuffled & \textbf{\boldmath +0.45 [0.32, 0.59]} & \textbf{\boldmath +0.50 [0.42, 0.57]} & \textbf{\boldmath +0.51 [0.42, 0.60]} \\
Statistics $\to$ + shuffled log-sig. & \textbf{\boldmath +0.37 [0.29, 0.45]} & \textbf{\boldmath +0.23 [0.16, 0.30]} & \textbf{\boldmath +0.30 [0.23, 0.37]} \\
Statistics $\to$ + shuffled raw sig. & \textbf{\boldmath +0.89 [0.71, 1.07]} & \textbf{\boldmath +0.34 [0.22, 0.46]} & \textbf{\boldmath +0.32 [0.25, 0.40]} \\
B3 $\to$ + shuffled log-sig. & $-0.02$ [$-0.06$, 0.02] & \textbf{\boldmath +0.06 [0.01, 0.10]} & +0.00 [$-0.01$, 0.01] \\
B3 $\to$ + shuffled raw sig. & \textbf{\boldmath +0.58 [0.45, 0.72]} & +0.07 [$-0.01$, 0.14] & $-0.02$ [$-0.08$, 0.04] \\
\addlinespace
\multicolumn{4}{@{}l}{\textit{Design B}} \\
Statistics + log-sig.: ordered $-$ shuffled & \textbf{\boldmath +1.77 [1.50, 2.03]} & \textbf{\boldmath +0.70 [0.48, 0.93]} & \textbf{\boldmath +1.16 [0.96, 1.36]} \\
Statistics + raw sig.: ordered $-$ shuffled & \textbf{\boldmath +2.95 [2.59, 3.32]} & \textbf{\boldmath +0.70 [0.52, 0.88]} & \textbf{\boldmath +1.06 [0.86, 1.27]} \\
B3 + log-sig.: ordered $-$ shuffled & \textbf{\boldmath +1.45 [1.23, 1.67]} & \textbf{\boldmath +0.32 [0.18, 0.45]} & \textbf{\boldmath +0.45 [0.30, 0.60]} \\
B3 + raw sig.: ordered $-$ shuffled & \textbf{\boldmath +2.50 [2.19, 2.81]} & \textbf{\boldmath +0.32 [0.14, 0.51]} & \textbf{\boldmath +0.55 [0.40, 0.70]} \\
Statistics $\to$ + shuffled log-sig. & +0.11 [$-0.02$, 0.25] & \textbf{\boldmath +0.45 [0.24, 0.67]} & \textbf{\boldmath +0.28 [0.13, 0.42]} \\
Statistics $\to$ + shuffled raw sig. & \textbf{\boldmath +1.58 [1.25, 1.90]} & \textbf{\boldmath +0.58 [0.31, 0.86]} & \textbf{\boldmath +0.39 [0.24, 0.55]} \\
B3 $\to$ + shuffled log-sig. & +0.01 [$-0.03$, 0.04] & +0.04 [$-0.11$, 0.18] & +0.02 [$-0.06$, 0.10] \\
B3 $\to$ + shuffled raw sig. & \textbf{\boldmath +1.56 [1.32, 1.80]} & +0.08 [$-0.12$, 0.28] & $-0.02$ [$-0.17$, 0.13] \\
\bottomrule
\end{tabular}
}
\end{table}

\subsection{Truncation order}

Table~\ref{tab:s-trunc} varies the log-signature truncation level (dimension in
parentheses) on the Statistics block, for a random forest of fixed depth and a
validation-tuned logistic regression. The noise control appends 32 independent
standard-normal columns, the dimension of the level-4 log-signature, to Statistics.
In Design A the gain is largest at level 2 and shrinks with further levels, at about the
rate of the noise control; in Design B each additional level adds information.

\begin{table}[htbp]
\centering
\caption{Balanced-accuracy gain by log-signature truncation level.}
\label{tab:s-trunc}
\resizebox{\textwidth}{!}{\begin{tabular}{@{}llcccccc@{}}
\toprule
 & & \multicolumn{3}{c}{Statistics + log-sig.\ minus Statistics} & \multicolumn{2}{c}{Step} & Noise \\
\cmidrule(lr){3-5}\cmidrule(lr){6-7}\cmidrule(l){8-8}
Design & Learner & Level 2 (6) & Level 3 (14) & Level 4 (32) & $2\to3$ & $3\to4$ & control \\
\midrule
Design A & RF & \textbf{\boldmath +0.39 [0.32, 0.46]} & \textbf{\boldmath +0.29 [0.22, 0.36]} & \textbf{\boldmath +0.21 [0.11, 0.30]} & \textbf{\boldmath $-0.10$ [$-0.14$, $-0.05$]} & \textbf{\boldmath $-0.08$ [$-0.14$, $-0.03$]} & \textbf{\boldmath $-0.17$ [$-0.26$, $-0.08$]} \\
Design A & Logit & \textbf{\boldmath +0.82 [0.71, 0.93]} & \textbf{\boldmath +0.83 [0.68, 0.97]} & \textbf{\boldmath +0.68 [0.54, 0.82]} & +0.01 [$-0.08$, 0.09] & \textbf{\boldmath $-0.15$ [$-0.20$, $-0.09$]} & \textbf{\boldmath $-0.07$ [$-0.11$, $-0.03$]} \\
Design B & RF & \textbf{\boldmath +0.51 [0.33, 0.68]} & \textbf{\boldmath +0.75 [0.53, 0.97]} & \textbf{\boldmath +1.11 [0.79, 1.43]} & \textbf{\boldmath +0.24 [0.13, 0.36]} & \textbf{\boldmath +0.36 [0.19, 0.53]} & +0.00 [$-0.20$, 0.20] \\
Design B & Logit & \textbf{\boldmath +0.74 [0.56, 0.91]} & \textbf{\boldmath +1.88 [1.59, 2.17]} & \textbf{\boldmath +2.22 [1.89, 2.55]} & \textbf{\boldmath +1.15 [0.94, 1.35]} & \textbf{\boldmath +0.34 [0.18, 0.50]} & $-0.04$ [$-0.10$, 0.03] \\
\bottomrule
\end{tabular}
}
\end{table}

\clearpage
\section{Detection results}\label{sec:s-detection}

\subsection{All detectors and false-alarm targets}

Tables~\ref{tab:s-det-a} and~\ref{tab:s-det-b} report the test results behind Figure~A.1 of
the main text. Operating points are selected on validation paths and applied once to test
paths. Chance DP is the detection probability of a memoryless alarm process firing at the
detector's realized alarm rate, $1-\exp\{-101\times\text{rate}\}$. Classifier detectors use
the validation-selected learner for each representation.

\begin{table}[htbp]
\centering
\caption{Detection at matched false-alarm targets, Design A.}
\label{tab:s-det-a}
\resizebox{\textwidth}{!}{\begin{tabular}{@{}lcccc@{}}
\toprule
Detector & Realized FA / 1,000 & DP (\%) [95\% CI] & Chance DP (\%) & Mean delay [95\% CI] \\
\midrule
\multicolumn{5}{@{}l}{\textit{Target: 0.5 false alarms per 1,000 observations}} \\
CUSUM & 0.48 & 12.4 [10.8, 14.0] & 6.9 & 59.8 [56.9, 62.7] \\
HMM filter & 0.51 & 4.9 [3.6, 6.1] & 5.9 & 62.8 [55.3, 70.4] \\
Statistics & 0.46 & 5.7 [4.5, 6.9] & 5.5 & 55.8 [48.4, 63.2] \\
Statistics + log-sig. & 0.49 & 5.5 [4.1, 7.0] & 5.8 & 55.2 [47.2, 63.1] \\
Statistics + raw sig. & 0.50 & 5.2 [4.0, 6.4] & 5.8 & 56.6 [49.5, 63.6] \\
B3 & 0.46 & 5.1 [4.0, 6.3] & 5.5 & 63.3 [56.5, 70.0] \\
B3 + log-sig. & 0.49 & 5.2 [3.9, 6.6] & 5.7 & 64.2 [57.1, 71.4] \\
B3 + raw sig. & 0.45 & 5.2 [4.0, 6.5] & 5.4 & 58.5 [51.4, 65.7] \\
\addlinespace
\multicolumn{5}{@{}l}{\textit{Target: 1 false alarms per 1,000 observations}} \\
CUSUM & 0.93 & 22.8 [20.8, 24.9] & 12.8 & 58.3 [55.5, 61.0] \\
HMM filter & 0.86 & 18.0 [16.3, 19.6] & 11.3 & 65.2 [62.7, 67.6] \\
Statistics & 0.86 & 18.2 [16.5, 19.9] & 11.4 & 64.7 [61.8, 67.5] \\
Statistics + log-sig. & 0.89 & 18.2 [16.4, 19.9] & 11.6 & 63.6 [61.2, 66.0] \\
Statistics + raw sig. & 0.87 & 17.5 [15.7, 19.2] & 11.4 & 62.9 [60.7, 65.2] \\
B3 & 0.85 & 17.3 [15.6, 19.0] & 11.1 & 64.6 [62.1, 67.0] \\
B3 + log-sig. & 0.88 & 18.3 [16.7, 20.0] & 11.6 & 64.6 [62.2, 67.1] \\
B3 + raw sig. & 0.90 & 17.5 [15.9, 19.2] & 11.6 & 64.9 [62.3, 67.5] \\
\addlinespace
\multicolumn{5}{@{}l}{\textit{Target: 2 false alarms per 1,000 observations}} \\
CUSUM & 1.74 & 35.0 [32.7, 37.3] & 21.4 & 56.7 [54.3, 59.1] \\
HMM filter & 1.98 & 46.5 [43.8, 49.3] & 24.9 & 60.1 [58.4, 61.8] \\
Statistics & 1.85 & 43.0 [39.4, 46.5] & 23.4 & 59.3 [57.6, 61.1] \\
Statistics + log-sig. & 1.85 & 42.3 [39.0, 45.6] & 23.2 & 59.4 [57.6, 61.2] \\
Statistics + raw sig. & 1.94 & 44.5 [41.5, 47.6] & 24.4 & 59.4 [57.6, 61.2] \\
B3 & 1.86 & 42.6 [39.3, 45.8] & 23.4 & 61.5 [59.7, 63.3] \\
B3 + log-sig. & 1.85 & 44.5 [41.5, 47.4] & 23.6 & 60.6 [58.8, 62.4] \\
B3 + raw sig. & 1.90 & 46.1 [43.1, 49.2] & 24.2 & 60.3 [58.6, 61.9] \\
\addlinespace
\multicolumn{5}{@{}l}{\textit{Target: 5 false alarms per 1,000 observations}} \\
CUSUM & 4.64 & 58.5 [56.3, 60.6] & 43.8 & 48.4 [46.8, 50.0] \\
HMM filter & 4.73 & 74.8 [72.5, 77.0] & 45.9 & 46.1 [44.8, 47.3] \\
Statistics & 4.71 & 75.9 [74.1, 77.7] & 46.0 & 49.4 [48.2, 50.6] \\
Statistics + log-sig. & 4.64 & 76.5 [74.6, 78.5] & 45.7 & 49.7 [48.6, 50.9] \\
Statistics + raw sig. & 4.66 & 77.6 [75.7, 79.6] & 45.9 & 48.6 [47.3, 49.9] \\
B3 & 4.68 & 76.3 [74.3, 78.4] & 45.9 & 50.0 [49.0, 51.0] \\
B3 + log-sig. & 4.66 & 77.0 [74.9, 79.0] & 45.8 & 50.0 [48.9, 51.1] \\
B3 + raw sig. & 4.59 & 76.7 [74.5, 78.8] & 45.4 & 49.6 [48.5, 50.7] \\
\bottomrule
\end{tabular}
}
\end{table}

\begin{table}[htbp]
\centering
\caption{Detection at matched false-alarm targets, Design B.}
\label{tab:s-det-b}
\resizebox{\textwidth}{!}{\begin{tabular}{@{}lcccc@{}}
\toprule
Detector & Realized FA / 1,000 & DP (\%) [95\% CI] & Chance DP (\%) & Mean delay [95\% CI] \\
\midrule
\multicolumn{5}{@{}l}{\textit{Target: 0.5 false alarms per 1,000 observations}} \\
CUSUM & 0.42 & 3.8 [3.1, 4.5] & 4.8 & 48.7 [42.1, 55.3] \\
HMM filter & 0.46 & 5.9 [4.8, 7.0] & 5.6 & 50.1 [44.9, 55.3] \\
Statistics & 0.44 & 6.2 [4.9, 7.4] & 5.5 & 51.8 [46.8, 56.8] \\
Statistics + log-sig. & 0.45 & 6.1 [5.1, 7.1] & 5.5 & 49.1 [43.7, 54.6] \\
Statistics + raw sig. & 0.47 & 6.2 [5.2, 7.2] & 5.7 & 50.6 [45.5, 55.7] \\
B3 & 0.44 & 6.4 [5.2, 7.6] & 5.5 & 49.9 [44.4, 55.4] \\
B3 + log-sig. & 0.47 & 6.3 [5.2, 7.3] & 5.7 & 51.1 [45.9, 56.4] \\
B3 + raw sig. & 0.45 & 6.1 [4.9, 7.3] & 5.5 & 50.8 [45.9, 55.7] \\
\addlinespace
\multicolumn{5}{@{}l}{\textit{Target: 1 false alarms per 1,000 observations}} \\
CUSUM & 0.85 & 8.8 [7.4, 10.2] & 9.6 & 47.3 [43.1, 51.5] \\
HMM filter & 0.89 & 10.8 [9.4, 12.3] & 10.4 & 48.7 [44.4, 53.0] \\
Statistics & 0.91 & 12.2 [10.7, 13.7] & 10.8 & 49.3 [45.6, 53.0] \\
Statistics + log-sig. & 0.87 & 11.1 [9.9, 12.3] & 10.3 & 51.6 [47.4, 55.8] \\
Statistics + raw sig. & 0.92 & 12.4 [10.8, 14.1] & 11.0 & 50.6 [47.1, 54.0] \\
B3 & 0.92 & 11.4 [9.9, 12.9] & 10.8 & 45.4 [41.3, 49.6] \\
B3 + log-sig. & 0.88 & 11.7 [10.2, 13.1] & 10.4 & 49.2 [45.4, 53.0] \\
B3 + raw sig. & 0.90 & 12.0 [10.2, 13.8] & 10.7 & 50.4 [46.2, 54.6] \\
\addlinespace
\multicolumn{5}{@{}l}{\textit{Target: 2 false alarms per 1,000 observations}} \\
CUSUM & 1.89 & 18.2 [15.9, 20.4] & 20.0 & 44.9 [41.7, 48.1] \\
HMM filter & 1.81 & 20.5 [18.6, 22.4] & 19.7 & 48.0 [44.4, 51.6] \\
Statistics & 1.81 & 22.5 [20.5, 24.5] & 20.1 & 48.0 [45.4, 50.6] \\
Statistics + log-sig. & 1.88 & 22.8 [20.8, 24.8] & 20.7 & 48.5 [45.9, 51.0] \\
Statistics + raw sig. & 1.86 & 24.8 [23.0, 26.7] & 20.8 & 49.3 [46.6, 52.0] \\
B3 & 1.87 & 21.5 [19.7, 23.3] & 20.4 & 48.3 [45.8, 50.8] \\
B3 + log-sig. & 1.88 & 23.2 [20.8, 25.5] & 20.8 & 46.5 [43.9, 49.1] \\
B3 + raw sig. & 1.82 & 23.5 [21.4, 25.6] & 20.3 & 52.0 [48.9, 55.2] \\
\addlinespace
\multicolumn{5}{@{}l}{\textit{Target: 5 false alarms per 1,000 observations}} \\
CUSUM & 4.47 & 39.8 [37.4, 42.3] & 40.6 & 42.8 [40.7, 45.0] \\
HMM filter & 4.66 & 47.0 [45.2, 48.9] & 42.7 & 45.6 [43.7, 47.5] \\
Statistics & 4.57 & 44.8 [42.3, 47.3] & 41.9 & 44.8 [42.6, 47.0] \\
Statistics + log-sig. & 4.69 & 45.3 [42.8, 47.9] & 42.7 & 46.4 [44.6, 48.1] \\
Statistics + raw sig. & 4.64 & 48.0 [45.2, 50.9] & 42.6 & 45.5 [43.7, 47.4] \\
B3 & 4.71 & 44.0 [41.7, 46.3] & 42.7 & 45.1 [43.3, 46.9] \\
B3 + log-sig. & 4.56 & 43.7 [41.4, 46.0] & 41.7 & 44.6 [42.8, 46.4] \\
B3 + raw sig. & 4.65 & 47.2 [44.8, 49.6] & 42.6 & 45.4 [43.7, 47.2] \\
\bottomrule
\end{tabular}
}
\end{table}

\subsection{Accuracy by time since the last change}

Table~\ref{tab:s-age} stratifies test windows by the number of observations since the
last true regime change and reports the gain of Statistics plus level-3 log-signatures
over Statistics, for a random forest of fixed depth and a validation-tuned logistic
regression.

\begin{table}[htbp]
\centering
\caption{Balanced-accuracy gain from log-signatures by time since the last change.}
\label{tab:s-age}
\small
\resizebox{\textwidth}{!}{\begin{tabular}{@{}lcccc@{}}
\toprule
 & \multicolumn{2}{c}{Design A} & \multicolumn{2}{c}{Design B} \\
\cmidrule(lr){2-3}\cmidrule(l){4-5}
Observations since change & RF & Logit & RF & Logit \\
\midrule
0--24 & $-0.08$ [$-0.26$, 0.11] & $-0.20$ [$-0.65$, 0.25] & $-0.31$ [$-0.86$, 0.24] & \textbf{\boldmath $-1.32$ [$-2.47$, $-0.17$]} \\
25--49 & +0.13 [$-0.15$, 0.40] & \textbf{\boldmath +2.86 [2.32, 3.40]} & \textbf{\boldmath +0.75 [0.09, 1.40]} & \textbf{\boldmath +1.28 [0.41, 2.14]} \\
50--99 & +0.19 [$-0.07$, 0.45] & \textbf{\boldmath +2.86 [2.28, 3.43]} & \textbf{\boldmath +1.31 [0.83, 1.78]} & \textbf{\boldmath +2.68 [1.84, 3.52]} \\
100+ & \textbf{\boldmath +0.34 [0.27, 0.42]} & \textbf{\boldmath +0.55 [0.38, 0.73]} & \textbf{\boldmath +0.74 [0.50, 0.97]} & \textbf{\boldmath +2.00 [1.65, 2.35]} \\
\bottomrule
\end{tabular}
}
\end{table}

\clearpage
\section{Forecasting study}\label{sec:s-market}

\subsection{Selected learners}

Table~\ref{tab:s-market-learners} counts, over the 21 annual refits from 2005 to 2025,
how often each learner is selected on validation balanced accuracy.

\begin{table}[htbp]
\centering
\caption{Validation-selected learner by representation, number of test years.}
\label{tab:s-market-learners}
\small
\begin{tabular}{@{}lccc@{}}
\toprule
Representation & Logit & RF & GBM \\
\midrule
\multicolumn{4}{@{}l}{\textit{United States}} \\
HMM probability block & 5 & 10 & 6 \\
Statistics & 12 & 4 & 5 \\
Statistics + log-sig. & 13 & 5 & 3 \\
Statistics + raw sig. & 13 & 6 & 2 \\
B3 & 7 & 10 & 4 \\
B3 + log-sig. & 10 & 9 & 2 \\
B3 + raw sig. & 8 & 11 & 2 \\
\addlinespace
\multicolumn{4}{@{}l}{\textit{Developed ex US}} \\
HMM probability block & 8 & 10 & 3 \\
Statistics & 11 & 9 & 1 \\
Statistics + log-sig. & 11 & 7 & 3 \\
Statistics + raw sig. & 11 & 7 & 3 \\
B3 & 6 & 4 & 11 \\
B3 + log-sig. & 7 & 8 & 6 \\
B3 + raw sig. & 6 & 5 & 10 \\
\bottomrule
\end{tabular}

\end{table}

\subsection{B3 relative to Statistics by year}

Figure~1 of the main text compares the rich and compact conventional baselines year by
year, each with its validation-selected learner. B3 is not more variable than Statistics: the
standard deviation of yearly balanced accuracy is 18.4 against 19.5 points in the United
States and 11.4 against 10.4 in developed markets excluding the United States, and a
Pitman--Morgan test of equal variances does not reject in either market ($p=0.45$ and
0.52). What differs across markets is the level of B3. In the United States B3 falls
below Statistics by 3.7 points on average [$-6.6$, $-0.8$] and in 14 of 21 years; outside
the United States it lies 1.6 points above [$-1.5$, 4.8] and falls below in 8 years. The
year-by-year difference between the two markets, $-5.3$ points, is resolved ($p=0.009$,
paired over years). In the United States the gain from adding raw signatures to B3 is
largest in the years in which B3 falls furthest below Statistics (correlation $-0.75$);
outside the United States the relation is weak ($-0.24$). Years are treated as
independent in these tests, and yearly balanced accuracy is noisy in years dominated by
one volatility tercile (2017 and 2022 in the United States).

\subsection{Remaining signature rules}

Table~\ref{tab:s-remaining} reports the volatility-managed results for the two signature
rules that were not prespecified, Statistics plus raw signatures and B3 plus
log-signatures, in the format of Table~3 of the main text. They are computed from the stored
out-of-sample weights with the same Sharpe-ratio test, by
\path{market_economics_all_rules.py}, which reproduces every row of Table~3 exactly. Mean and
volatility are annualized and in percent, Max DD is the maximum drawdown in percent, Turnover
is the average daily absolute change in exposure, and $p$ tests equality with the Sharpe
ratio of the Statistics rule.

In the United States neither rule differs detectably from the Statistics rule ($p=0.21$ and
0.76). In developed markets excluding the United States, Statistics plus raw signatures
attains a Sharpe ratio of 0.543 against 0.410 ($p=0.016$), the only nominally resolved
Sharpe-ratio difference in the study, although the same representation is not more accurate
than Statistics in that market (Table~1 of the main text). The rule was not prespecified, and
across the ten Sharpe-ratio comparisons of the two markets its Holm-adjusted $p$-value is
0.16, so the difference is not resolved after correction for multiplicity.

\begin{table}[htbp]
\centering
\caption{Volatility-managed market exposure for the remaining signature rules, 2005--2025.}
\label{tab:s-remaining}
\small
\begin{tabular}{@{}lcccccc@{}}
\toprule
Exposure rule & Mean & Vol. & Sharpe & Max DD & Turnover & $p$ \\
\midrule
\multicolumn{7}{@{}l}{\textit{Panel A. United States}} \\
Statistics & 9.53 & 16.85 & 0.566 & $-33.7$ & 0.067 & -- \\
Statistics + raw sig. & 10.82 & 17.33 & 0.625 & $-36.9$ & 0.060 & 0.208 \\
B3 + log-sig. & 9.78 & 16.84 & 0.581 & $-36.4$ & 0.064 & 0.763 \\
\addlinespace
\multicolumn{7}{@{}l}{\textit{Panel B. Developed markets excluding the United States}} \\
Statistics & 6.23 & 15.21 & 0.410 & $-46.6$ & 0.055 & -- \\
Statistics + raw sig. & 8.91 & 16.42 & 0.543 & $-45.8$ & 0.059 & 0.016 \\
B3 + log-sig. & 7.09 & 15.00 & 0.473 & $-43.3$ & 0.071 & 0.189 \\
\bottomrule
\end{tabular}

\end{table}

\subsection{All forecast scores}

Table~\ref{tab:s-scores} reports every forecast of the main text under four measures:
balanced accuracy (BA), the ranked probability score (RPS), the logarithmic score and the
Brier score, the last two strictly proper (Gneiting and Raftery, 2007). The RPS and the Brier
score are multiplied by 100 and the logarithmic score is in nats; lower scores are better, and
bold marks the best value in each column. Persistence issues one-hot forecasts, so its
logarithmic score is infinite.

The three scores agree in direction for every contrast of the main text: for each of the
eight signature augmentations of the classifiers and the six augmentations of the HAR
regression, the RPS, the logarithmic score and the Brier score all improve or all deteriorate.
Their resolution after Holm correction differs in three cases. The logarithmic score, which
penalizes confident errors most heavily, resolves a deterioration from adding log-signatures
to Statistics outside the United States (Holm-adjusted $p=0.044$) that the RPS does not
($p=0.53$). Conversely, it resolves the gains from adding log-signatures to Statistics and raw
signatures to the HAR regression in the United States only nominally (Holm-adjusted $p=0.074$
and 0.093), whereas the RPS and the Brier score resolve both.

\begin{table}[htbp]
\centering
\caption{Out-of-sample forecasts under all four measures, 2005--2025.}
\label{tab:s-scores}
\resizebox{\textwidth}{!}{\begin{tabular}{@{}lcccccccc@{}}
\toprule
 & \multicolumn{4}{c}{United States} & \multicolumn{4}{c}{Developed ex US} \\
\cmidrule(lr){2-5}\cmidrule(l){6-9}
Forecast & BA (\%) & RPS & Log & Brier & BA (\%) & RPS & Log & Brier \\
\midrule
\multicolumn{9}{@{}l}{\textit{Benchmarks}} \\
Persistence & 55.81 & 25.29 & -- & 87.95 & 56.82 & 23.96 & -- & 83.45 \\
Transition & 55.81 & 17.53 & 0.957 & 56.81 & 56.82 & 17.27 & 0.921 & 54.94 \\
HAR & 56.22 & 16.35 & 0.902 & 53.47 & \textbf{57.51} & 16.04 & 0.867 & 51.79 \\
\addlinespace
\multicolumn{9}{@{}l}{\textit{HAR augmented}} \\
HAR + Statistics & 55.77 & 16.17 & 0.894 & 53.07 & 56.19 & 15.92 & 0.858 & 51.48 \\
HAR + log-sig. & 58.49 & 15.53 & 0.877 & 51.41 & 57.19 & \textbf{15.57} & \textbf{0.846} & \textbf{50.73} \\
HAR + raw sig. & \textbf{58.92} & \textbf{15.33} & \textbf{0.866} & \textbf{50.87} & 56.17 & 15.72 & 0.855 & 50.97 \\
\addlinespace
\multicolumn{9}{@{}l}{\textit{Classifiers}} \\
HMM probability block & 56.10 & 16.58 & 0.941 & 54.82 & 56.00 & 16.47 & 0.894 & 53.25 \\
Statistics & 54.25 & 18.22 & 1.163 & 60.86 & 54.77 & 15.88 & 0.876 & 51.99 \\
Statistics + log-sig. & 56.38 & 16.70 & 1.005 & 55.79 & 56.09 & 16.40 & 0.939 & 53.70 \\
Statistics + raw sig. & 57.92 & 16.50 & 0.966 & 55.11 & 54.28 & 16.74 & 1.108 & 55.77 \\
B3 & 50.26 & 18.18 & 1.041 & 60.18 & 55.59 & 17.39 & 1.018 & 57.24 \\
B3 + log-sig. & 53.87 & 17.14 & 0.974 & 57.44 & 54.46 & 17.03 & 0.957 & 56.35 \\
B3 + raw sig. & 53.98 & 18.18 & 1.128 & 60.91 & 52.40 & 17.92 & 1.106 & 59.76 \\
\bottomrule
\end{tabular}
}
\end{table}

\clearpage
\subsection{HAR augmentation tests}

Table~\ref{tab:s-har-pvals} reports tests of the three HAR augmentations against plain
HAR in each market under all four forecast scores. The six comparisons (two markets
and three augmentations) form a separate Holm family for each score. Balanced-accuracy
$p$-values come from two-sided stationary-bootstrap tests with 2,000 resamples and
a mean block length of 21 days. Score $p$-values come from two-sided
Diebold--Mariano tests with a Bartlett HAC variance and bandwidth 21. The adjusted
$p$-values are computed from the unrounded values; bold denotes an adjusted
$p<0.05$.

\begin{table}[htbp]
\centering
\caption{HAR augmentation versus plain HAR: raw and Holm-adjusted $p$-values, 2005--2025.}
\label{tab:s-har-pvals}
\small
\begin{tabular}{@{}llcc@{}}
\toprule
Augmentation & Score & Raw $p$ & Holm-adjusted $p$ \\
\midrule
\multicolumn{4}{@{}l}{\textit{United States}} \\
HAR + Statistics & BA & 0.6240 & 1.0000 \\
 & RPS & 0.4351 & 0.8703 \\
 & Log score & 0.4296 & 1.0000 \\
 & Brier score & 0.5006 & 1.0000 \\
\addlinespace
HAR + log-sig. & BA & 0.0165 & 0.0825 \\
 & RPS & 0.0201 & 0.0972 \\
 & Log score & 0.1956 & 0.7824 \\
 & Brier score & 0.0129 & 0.0643 \\
\addlinespace
HAR + raw sig. & BA & 0.0065 & \textbf{0.0390} \\
 & RPS & 0.0008 & \textbf{0.0045} \\
 & Log score & 0.0155 & 0.0930 \\
 & Brier score & 0.0010 & \textbf{0.0059} \\
\addlinespace
\multicolumn{4}{@{}l}{\textit{Developed ex US}} \\
HAR + Statistics & BA & 0.1160 & 0.4640 \\
 & RPS & 0.5708 & 0.8703 \\
 & Log score & 0.4047 & 1.0000 \\
 & Brier score & 0.5650 & 1.0000 \\
\addlinespace
HAR + log-sig. & BA & 0.7070 & 1.0000 \\
 & RPS & 0.0194 & 0.0972 \\
 & Log score & 0.0241 & 0.1207 \\
 & Brier score & 0.0345 & 0.1381 \\
\addlinespace
HAR + raw sig. & BA & 0.2775 & 0.8325 \\
 & RPS & 0.2656 & 0.7968 \\
 & Log score & 0.3976 & 1.0000 \\
 & Brier score & 0.2485 & 0.7455 \\
\bottomrule
\end{tabular}

\end{table}